\documentclass[review]{elsarticle}

\journal{Journal of \LaTeX\ Templates}

\usepackage{graphicx}
\usepackage{float}
\usepackage{caption}
\usepackage{subfigure} 
\usepackage{url}
\usepackage{amsmath}
\usepackage{amssymb}
\usepackage[usenames,dvipsnames]{xcolor}
\usepackage{color}
\usepackage{setspace}
\usepackage{supertabular}
\usepackage{xspace}
\usepackage{soul}
\soulregister\cite7
\soulregister\ref7
\usepackage{balance}
\usepackage{amsfonts}
\usepackage[noend]{algpseudocode}
\usepackage{algorithm}
\usepackage{algorithmicx}
\usepackage{pifont}
\usepackage{epstopdf}
\usepackage{bm}
\usepackage{multicol}
\usepackage{multirow}
\usepackage{tabularx}
\usepackage{makecell}
\usepackage{array}
\newcolumntype{C}[1]{>{\vspace{0.2em}\begin{minipage}{#1}\centering\let\newline\\
\arraybackslash\hspace{0pt}}m{#1}<{\end{minipage}\vspace{0.2em}}}
\usepackage{booktabs}

\usepackage{setspace}

\begin{document}

\captionsetup[figure]{labelfont={bf},labelformat={default},labelsep=period,name={Fig.}}

\begin{frontmatter}

\title{Energy Trading in Microgrids for Synergies among Electricity, Hydrogen and Heat Networks\tnoteref{mytitlenote}}
\tnotetext[mytitlenote]{This work was supported by the National Key Research and Development Program of China (Grant No.2016YFB0901900), and in part by the NSF of China (Grants No. 61731012, 61573245, 61922058, and 61973264).}


\author[mymainaddress,mysecondaryaddress]{Dafeng Zhu}

\author[mymainaddress,mysecondaryaddress]{Bo Yang\corref{mycorrespondingauthor}}
\cortext[mycorrespondingauthor]{Corresponding author}
\ead{bo.yang@sjtu.edu.cn}

\author[mymainaddress,mysecondaryaddress]{Qi Liu}
\author[mythirdaddress]{Kai Ma}
\author[mymainaddress,mysecondaryaddress]{Shanying Zhu}
\author[myfourthaddress]{Chengbin Ma}
\author[mymainaddress,mysecondaryaddress]{Xinping Guan}

\address[mymainaddress]{Department of Automation, Shanghai Jiao Tong University, Shanghai 200240, China}
\address[mysecondaryaddress]{Key Laboratory of System Control and Information Processing, Ministry of Education of China, Shanghai 200240, China}
\address[mythirdaddress]{Key Laboratory of Industrial Computer Control Engineering of Hebei Province, Yanshan University, Qinhuangdao 066004, China}
\address[myfourthaddress]{University of Michigan-Shanghai Jiao Tong University Joint Institute, Shanghai Jiao Tong University, Shanghai 200240, China}


\begin{abstract}
{
{The emerging paradigm of interconnected microgrids advocates energy trading or sharing among multiple microgrids. It helps make full use of the temporal availability of energy and diversity in operational costs when meeting various energy loads.}}
 However, energy trading might not completely absorb excess renewable energy. A multi-energy management framework including fuel cell vehicles, energy storage, {
 {combined heat and power system}}, and renewable energy is proposed, and the characteristics and scheduling arrangements of fuel cell vehicles are considered to further improve the local absorption of the renewable energy and enhance the economic benefits of microgrids. {
 {While intensive research has been conducted on energy scheduling and trading problem, a fundamental question still remains unanswered on microgrid economics. Namely, due to multi-energy coupling, stochastic renewable energy generation and demands, when and how a microgrid should schedule and trade energy with others, which maximizes its long-term benefit.  This paper designs a joint energy scheduling and trading algorithm based on Lyapunov optimization and a double-auction mechanism. 
 Its purpose is to determine the valuations of energy in the auction, optimally schedule energy distribution, and strategically purchase and sell energy with the current electricity prices. 
Simulations based on real data show that each individual microgrid, under the management of the proposed algorithm, can achieve a time-averaged profit that is arbitrarily close to an optimum value, while avoiding compromising its own comfort.}}

\end{abstract}

\begin{keyword}
\texttt Multi-energy microgrid \sep energy trading \sep energy storage \sep Lyapunov optimization\sep double-auction  
\end{keyword}

\end{frontmatter}


\section{Introduction}
Traditional power grids consume fossil fuels to generate electricity and transmit electricity over long distances, which results in quick depletion of fossil fuel resources and serious environmental pollution. This motivates the study of distributed microgrids (MGs), which can efficiently realize investment deferral \cite{ArmendCoordinated}, local balance \cite{ZhangPeer}, resiliency advancement \cite{RenEnabling}, security reinforcement \cite{Zhu2015Microgrid} and reduce greenhouse gas emissions and energy losses by using renewable energy sources {
{\cite{ZhangCredit}}}. However, renewable energy generation is stochastic, which may influence energy reliability and quality. 
Meanwhile, considering the heat demands of users, the combined heat and power (CHP) system, which can efficiently generate both electricity and heat simultaneously by consuming natural gas, is introduced. 
Energy storage also plays a key role in improving energy reliability  by storing extra energy to be used in the future.
However, it is not efficient and economic for individual MG to serve its users because of the mismatch between renewable energy generation and electricity demand. 

Geographically distributed MGs can improve energy reliability and efficiency by sharing energy. However, the MG is selfish and wants to minimize its own cost for sharing energy. Only if its benefit cannot be lowered can a MG be incentivized to join energy trading. This needs an effective method to carry out energy scheduling and trading among multiple MGs in order to achieve benefit maximization for individual MGs. Several inter-related decisions are involved: (1) Energy pricing: what method should be adopted for energy sale and purchase among multiple MGs, and at what prices? (2) Energy scheduling: with time-varying demand and renewable generation of each MG, should a MG serve the demand using its own energy storage or trading with other MGs? When local energy storage and energy trading cannot satisfy the demand, should a MG serve the demand by purchasing energy from utility companies or CHP system, and is it necessary to exploit time-varying electricity prices? These decisions should be optimally and efficiently made online while guaranteeing individual MG's benefits for a long period.
Therefore, a joint algorithm for energy scheduling and trading for MGs is designed. 
A double-auction mechanism is proposed to determine the purchase price and selling price, increase the economic benefits of MGs, and ensure the truthfulness of the information that MGs submit in energy trading. 

However, owing to the limitation of the battery storage, the MG might not fully exploit the time-diversity of renewable energy generation. In order to improve the utilization of renewable energy generation, we can introduce hydrogen into the MG and use excess renewable energy to electrolyze water to produce and store hydrogen in hydrogen storage tanks. 
Fuel cell vehicles can convert hydrogen into electricity to supply the MG's energy demand when the MG is short of energy, and fuel cell vehicles can be used for transportation. The following few advantages contain the reasons we introduce hydrogen: First, for the same size of energy storage, hydrogen storage can provide larger amounts of energy than batteries and can be filled in a few minutes. A number of facilities that integrate renewable energy and energy storage are under operation all over the world and most of them use hydrogen for energy storage in both stand-alone and grid-tied power generation systems 
{\cite{KyriakarakosPolygeneration}}. In these facilities, the hydrogen storage system is often coupled with a battery bank for short-term energy storage, thus achieving a hybrid poly-generation system. Proper integration of hydrogen storage systems and batteries increases bus stability and enhances the management of intermittent power peaks and transient loads \cite{Little2007Electrical}. Second, the entire electricity-hydrogen conversion process only utilizes water and is carbon free. Last but not least, hydrogen can be purchased from a hydrogen-producing company and used for the transportation of fuel cell vehicles. Fuel cell vehicles are particularly suited to provide spinning reserves and peak power to the grid {
{\cite{Lipman2004Fuel}}}.  In contrast to plug-in electric vehicles, fuel cell vehicles can be operated continuously and have very low emissions \cite{Lipman2004Fuel}. Hydrogen, as a clean energy with high calorific value, is attracting wide attention. Therefore, the car as a power plant (CaPP) \cite{Wijk2014Our} is presented to introduce a controllable energy system, which uses fuel cell vehicles as dispatchable power plants \cite{Fernandes2016Fuel}.
Considering that the average driving time of vehicles is less than 10\% of the whole day, vehicles can generate electricity by combusting hydrogen in a cleaner way than other power systems when they are parked, and there is a huge potential for fuel cell vehicles to take replace traditional power plants or reduce the number of new plants in the future. 
Therefore, the synergies between hydrogen and electricity can be explored to increase the benefits of MGs. 

 In particular, the main contributions of this paper are as follows:

\begin{itemize}
\item 
A multi-energy management framework that includes fuel cell vehicles, energy storage, CHP system, and renewable energy is proposed. The synergies between hydrogen and electricity can further improve the local absorption of the excess renewable energy and the economic benefits of MGs.
\item 
A joint energy scheduling and trading algorithm based on Lyapunov optimization and a double-auction mechanism is designed to optimize the long-term energy cost of each MG.
\item Through theoretical analysis, the proposed algorithm can achieve better trade-off among energy trading cost, energy storage and users' satisfaction. Moreover, by using practical data sets, the effectiveness of the proposed algorithm is verified.
\end{itemize}

In the rest of the paper, Section II introduces related works. Section III describes the system model and cost functions. Then, Section IV proposes a joint algorithm based on Lyapunov optimization and a double-auction mechanism for the energy scheduling and trading problem, and proves the theoretical performance of this algorithm. Section V shows the numerical results, and Section VI concludes the paper. 
 
 \section{Related Works}
 Energy sharing is a way to reduce the imbalance of supply and demand of MGs and improve the local consumption of renewable energy.
A number of research efforts have been conducted. In \cite{2}, it is shown that energy sharing allows participants to exchange energy in order to lower reliance on the utility company. In \cite{3}, the authors demonstrate that the development of peer-to-peer energy sharing has a significant advantage to prosumers in both earning revenues and reducing energy costs. In \cite{4} because of stochastic renewable energy generation, nanogrids form a nanogrid cluster that shares renewable energy. In \cite{5}, a real-time demand response model is presented to assist the energy sharing provider, which realizes the maximization of the energy sharing provider's utility.{
{ In \cite{Chen2018Analyzing}, energy trading and sharing schemes for multiple energy hubs are proposed to increase system flexibility and reduce the cost of the system. }}

However, owing to the randomness of renewable energy, it is difficult to schedule renewable energy sharing among multiple MGs and investigate the economic aspect. There are two types of market-based models that are applicable for resource management of energy sharing. The first one is the market model where resource owners decide the price based on users' demands by the game approach. 
For the first situation, two different models are proposed:
1) the prosumers decide the price of energy together \cite{6,8,Chen2019An}; 2) a leader--follower structure decides the price \cite{7,9,Motalleb2019Networked}. 
Liu et al. \cite{6} formulate a dynamical internal pricing model for the energy sharing of prosumers who decide the price of  energy. Lu et al. \cite{8} establish an informative game vector to perform price-based energy interactions among multiple MGs that decide the price. {
{Chen et al. \cite{Chen2019An} propose a novel energy sharing game for prosumers to determine the role of the buyer or seller and the sharing price.}}
Liu et al. \cite{7} propose a Stackelberg game approach, in which the MG operator acts as the leader and prosumers act as followers to decide the price together. Tushar et al. \cite{9} formulate a non-cooperative Stackelberg game to capture the interaction between the shared facility controller and the residential units that decide the price of energy to minimize the cost. Motalleb et al. \cite{Motalleb2019Networked} propose a networked Stackelberg  competition among firms to determine their optimal bids for the price of a market transaction.
The second one is the auction model, where every player acts independently and agrees privately on the price. 
According to the type of interactions between buyers and sellers, auctions can be divided into two classes: one-side auction \cite{11} and two-side auction {
{\cite{10}}}. The auction mechanism helps players benefit from cooperation and energy trading with little global information. The auction mechanism can make every player autonomously share the energy and automatically guarantee the truthfulness of the energy information. Therefore, an auction mechanism is used to determine the price of energy sharing in this study.

Energy storage and CaPP are also effective ways to reduce the imbalance of supply and demand in MGs, and improve the local consumption of renewable energy. 
 In \cite{Huang2013Optimal}, Huang et al. develop a low-complexity algorithm with energy storage management to minimize the average cost of a power-consuming entity. In \cite{Gatzianas2010Control}, Gatzianas et al. explicitly take the actual energy storage into account and construct an algorithm for energy management based on the Lyapunov optimization technique.  In \cite{Gayme2011Optimal}, Gayme et al. investigate distributed energy storages and illustrate their effects using an example along with time-varying demand profiles. {
 {In \cite{Good2019A}, Good et al. propose an aggregation modeling method for multi-energy conversion, storage, and demand to take advantage of distributed energy flexibility and provide multiple services.}}

The scheduling of vehicles and electrolyzers are the main aspects to be considered in the operational control of the CaPP. Centralized optimization approaches, such as minimizing operating costs \cite{Battistelli2013Generalized} or power losses \cite{Khodr2012Intelligent}, are used to address the scheduling problem of vehicles. In \cite{Shinoda2016Optimization}, the scheduling problem in the MG among renewable energy sources (RES), electrolyzer, and vehicle-to-grid (V2G) power is to minimize the power purchased from the grid. {
{In \cite{Jaramillo2016Optimal}, a multi-objective mixed-integer linear programming model is proposed for the scheduling in a grid-connected MG, and the startup constraints of the alkaline electrolyzer are explicitly modeled.}} 
In \cite{Chiesa2011Dynamic}, the electrolyzer levels out voltage fluctuations in a weak grid and improves the power quality of the MG based on a dynamic electrolyzer model.

{
{However, the existing works do not consider the coordinated operation and multi-energy demands of MGs after introducing hydrogen storage and fuel cell vehicles. 
In this paper, a multi-energy management framework including fuel cell vehicles, energy storage, CHP system, and renewable energy is proposed. The characteristics and scheduling arrangements of fuel cell vehicles are considered to further improve the local absorption of the renewable energy and enhance the economic benefits of MGs. This paper designs a joint energy scheduling and trading algorithm based on Lyapunov optimization and a double-auction mechanism.
 The dynamic and computationally efficient energy scheduling and trading algorithm is developed that determines the valuations of energy in the auction, optimally schedules energy distribution, and strategically purchases and sells energy with the current electricity prices. 
The implementation of the algorithm only depends on the current system states without the need of any priori information.
At the end, simulations based on real data are conducted to investigate the performance of the multi-energy management framework and demonstrate the effectiveness of the proposed algorithm.}}
 
\section{System Model}
This paper considers a system consisting of $n$ interconnected MGs, an electricity utility company, a gas utility company, and a hydrogen-producing company. Each MG is equipped with renewable energy, CHP system,  fuel cell vehicles, battery, hydrogen storage, boiler, and water tank, as shown in Fig.1. MGs can harvest renewable energy, such as wind and solar power. Fuel cell vehicles can generate electricity by consuming hydrogen. CHP system can consume natural gas to generate electricity, and at the same time, the generated heat follows its electricity production with fixed ratios. In addition, each MG can store extra energy for the demand in the future.

\begin{figure*}
\centering
\begin{minipage}[c]{1\textwidth}
\centerline{\includegraphics[width=\textwidth]{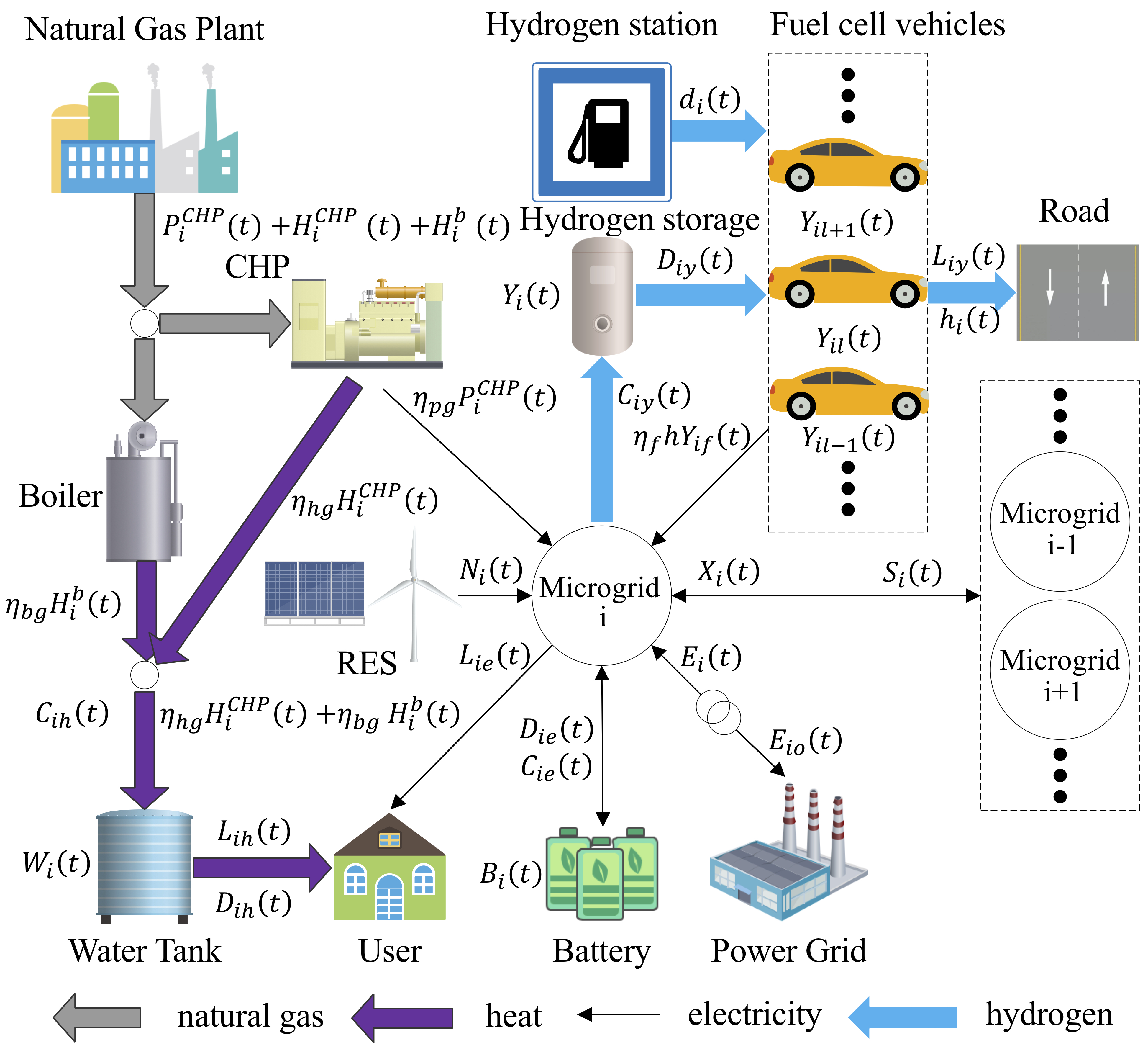}}
\end{minipage}
\caption{Energy flows of system.}
\label{fig1}
\end{figure*}

\subsection{Energy Purchase}
MG $i$ harvests $N_{i}(t)$ units of energy generated by renewable energy during one time slot. Here, one time slot is set to be one hour in order to coordinate with the simulation. The electricity utility company uses fossil energy to generate electricity, so it has huge energy generation at one time slot, which means that constraints on energy generation by the electricity utility company are not considered. The same assumption is applied to the gas utility company and the hydrogen-producing company. MG $i$ purchases $E_{i}(t)$ units of energy from the electricity utility company with price $p_{e}(t)$. From the gas utility company, MG $i$ purchases $P_{i}^{CHP}(t)$ and $H_{i}^{CHP}(t)$ units of gas to generate $\eta_{pg}P_{i}^{CHP}(t)$ units of electricity and $\eta_{hg}H_{i}^{CHP}(t)$ units of hot water by CHP system at time slot $t$. $\eta_{pg}$ and $\eta_{hg}$ are the conversion efficiency of CHP system from natural gas to electricity and heat, respectively. Moreover, MG $i$ purchases $H_{i}^{b}(t)$ units of gas to produce $\eta_{bg}H_{i}^{b}(t)$ units of hot water by boiler at time slot $t$. $\eta_{bg}$ is the conversion efficiency of boiler from natural gas to heat. The price of the gas is $p_{g}(t)$. When there is not enough hydrogen for fuel cell vehicles, MG $i$ will purchase $d_{i}(t)$ units of hydrogen from the hydrogen-producing company with price $p_{y}(t)$.

\subsection{Energy Demands} {\color{black}{MG $i$ needs to meet the electricity $L_{ie}(t)$, hydrogen $L_{iy}(t)$, and heat $L_{ih}(t)$ demands. Although these demands are stochastic, they still need to be met quickly and precisely.}}
\subsubsection{Electricity Demands} First, MG $i$ uses renewable energy to meet its users' electricity demands, $L_{ie}(t)$. If $N_{i}(t)>L_{ie}(t)$, extra renewable energy can be used for energy storage, water electrolysis, and energy trading. Otherwise, MG $i$ uses all renewable energy to serve its loads. The unsatisfied electricity loads are expressed as $L_{ie}(t)-N_{i}(t)$ and are served by the following methods:
\begin{itemize}
\item Discharge the battery. MG $i$ can draw $D_{ie}(t)$ units of electricity from the battery to serve unsatisfied electricity loads.
\item Generate electricity using hydrogen. Fuel cell vehicles can use hydrogen to generate $\eta_fhY_{if}(t)$ units of electricity.
\item Generate electricity using CHP system. CHP system can consume natural gas to generate $\eta_{pg}P_{i}^{CHP}(t)$ units of electricity to meet electricity demands.
\item Purchase electricity by energy trading. MG $i$ may acquire $X_{i}(t)$ units of electricity by trading with other MGs.
\item Purchase electricity from the electricity utility company. MG $i$ can purchase $E_{i}(t)$ units of electricity from the electricity utility company.
\end{itemize}
\subsubsection{Hydrogen Demands} First, vehicle $l_i$ uses $Y_{il}(t-1)$ units of hydrogen stored in the vehicle to meet its driving demands $h_{il}(t)$, which can be estimated from historical data. If $Y_{il}(t-1)>h_{il}(t)+Y_{il,min}$, the vehicle can drive normally. If $Y_{il}(t-1)\leq h_{il}(t)+Y_{il,min}$, the vehicle $l_i$ uses all hydrogen in the vehicle for driving. Deficient hydrogen is obtained from MG $i$ or purchased from a hydrogen-producing company. MG $i$ purchases $d_{i}(t-1)$ units of hydrogen to meet the total hydrogen demand $L_{iy}(t)$ of all vehicles at time slot $t$.
\subsubsection{Heat Demands} MG $i$ uses the hot water stored in the water tank to meet its heat demands. If these water cannot meet its heat demands, MG $i$ will use both CHP system and boiler to produce $\eta_{hg}H_{i}^{CHP}(t)+\eta_{bg}H_{i}^{b}(t)$ units of hot water to meet its heat demands $L_{ih}(t)$ at time slot $t$.

\subsection{Dynamic Model for Energy Storages}
Each MG has a battery that can store extra electricity generated by renewable energy generation, and a hot water tank to supply hot water. Meanwhile, 
 hydrogen storage is introduced, and the dynamic model for three types of energy storages is considered. For MG $i$, the electricity of battery, hydrogen of the storage, and equivalent thermal energy of the hot water tank are denoted by $B_{i}(t)$, $Y_{i}(t)$, and $W_{i}(t)$ at the end of one time slot, respectively. The electricity, hydrogen, and equivalent thermal energy are charged in the amounts of $C_{ie}(t)$, $C_{iy}(t)$, and $C_{ih}(t)$, and discharged in the amounts of  $D_{ie}(t)$, $D_{iy}(t)$, and $D_{ih}(t)$, respectively.
 Then, the energy storage dynamics can be obtained as:
\begin{equation}
B_{i}(t+1)=B_{i}(t)+C_{ie}(t)-D_{ie}(t)
\label{A1}
\end{equation}
\begin{equation}
Y_{i}(t+1)=Y_{i}(t)+C_{iy}(t)-D_{iy}(t)
\label{Y1}
\end{equation}
\begin{equation}
W_{i}(t+1)=W_{i}(t)+C_{ih}(t)-D_{ih}(t)
\label{A3}
\end{equation}
where $C_{iy}(t)$ denotes the amount of hydrogen injected into hydrogen storage, which is generated by the electrolyzer during one time slot. $\frac{hC_{iy}(t)}{\eta_e}$ is the energy consumed by the electrolyzer during one time slot, $\eta_e$ is the conversion efficiency of the electrolyzer from electricity to hydrogen, and $h$ is the heating value of hydrogen, which is {
{$1.4 \times 10^{8}$J/kg}}. Hydrogen needs to be compressed and stored. The compression energy is $c_1C_{iy}(t)$, and $c_1$ is the specific energy consumption of the compressor. 
To be specific, the operations of battery, hydrogen storage, and hot water tank of MG $i$ are subject to a lot of constraints. First, electricity charging and discharging will not happen simultaneously.
\begin{equation}
1_{C_{ie}(t)>0}+1_{D_{ie}(t)>0}\leq{1},
\label{cd1}
\end{equation}
\begin{center}
$1_{f(x)>0}=\left\{\begin{array}{cc}
1 & \mbox{if}\ f(x)>0\\
0 & \mbox{otherwise} \end{array}\right.$
\end{center}
Battery, hydrogen storage, and water tank of MG $i$ have finite capacities:
\begin{equation}
0\leq B_{i}(t) \leq B_{i,max}
\label{Bm}
\end{equation}
\begin{equation}
0\leq Y_{i}(t) \leq Y_{i,max}
\label{Ym}
\end{equation}
\begin{equation}
0\leq W_{i}(t) \leq W_{i,max}
\label{Wm}
\end{equation}
where $B_{i,max}$, $Y_{i,max}$ and $W_{i,max}$ are the upper bounds of the battery, hydrogen storage, and hot water tank's thermal energy. There are maximum electricity, hydrogen, and equivalent thermal energy charging $C_{ie,max}$, $C_{iy,max}$, $C_{ih,max}$ and discharging $D_{ie,max}$, $D_{iy,max}$, $D_{ih,max}$ during one time slot. Thus, the charging and discharging constraints of the energy storage are denoted by
\begin{equation}
0\leq C_{ie}(t) \leq C_{ie,max}, 0\leq D_{ie}(t) \leq D_{ie,max}
\label{Cem}
\end{equation}
\begin{equation}
0\leq C_{iy}(t) \leq C_{iy,max}, 0\leq D_{iy}(t) \leq D_{iy,max}
\label{Cym}
\end{equation}
\begin{equation}
0\leq C_{ih}(t) \leq C_{ih,max}, 0\leq D_{ih}(t) \leq D_{ih,max}
\label{Chm}
\end{equation}

The feasible control decision on $C_{ie}(t)$ and $D_{ie}(t)$ should meet constraints (\ref{cd1}), (\ref{Bm}), and (\ref{Cem}), simultaneously. {\color{black}{Since electricity charging and discharging will not happen simultaneously, the energy level of the battery cannot exceed the capacity of the battery, which means that $B_{i}(t)+C_{ie}(t)\leq B_{i,max}$. Meanwhile, the energy level of the battery cannot be negative, which means $B_{i}(t)-D_{ie}(t) \geq 0$. }}Therefore, the charging and discharging constraints of the battery are denoted as:
\begin{equation}
0\leq{C_{ie}(t)}\leq{\min[B_{i,max}-B_{i}(t),C_{ie,max}]}
\label{C1}
\end{equation}
\begin{equation}
0\leq{D_{ie}(t)}\leq{\min[B_{i}(t),D_{ie,max}]}
\label{C2}
\end{equation}

\subsection{Dynamic Model for Fuel Cell Vehicles}
Because fuel cell vehicles can act as controllable power plants, fuel cell vehicles are introduced and a dynamic model for fuel cell vehicles is considered. The model includes the transportation features and power generation of fuel cell vehicles. The transportation features are information about the departure, arrival time, and driving distance of each vehicle, which can be estimated. Power generation is determined by the transportation features and hydrogen storage of vehicles. The hydrogen in the vehicle $l_i$ is $Y_{il}(t)$ at the end of one time slot. The number of vehicles in MG $i$ is $L_{i}$. Then, the model of fuel cell vehicle $l_i$ is as follows:
\begin{equation}
Y_{il}(t+1)=\left\{\begin{array}{cc}
Y_{il}(t)+D_{iyl}(t)+d_{il}(t) & \mbox{injection} \\
Y_{il}(t)-Y_{ifl}(t) & \mbox{generation} \\
Y_{il}(t)-h_{il}(t) & \mbox{driving} \end{array}\right.
\label{Yc}
\end{equation}
\begin{equation}
\begin{split}
\sum_{l=1}^{L_i}D_{iyl}(t)=D_{iy}(t);
\sum_{l=1}^{L_i}d_{il}(t)=d_{i}(t)\\
\sum_{l=1}^{L_i}Y_{ifl}(t)=Y_{if}(t);
\sum_{l=1}^{L_i}h_{il}(t)=h_{i}(t)\\
\end{split}
\label{DdYh}
\end{equation}
\begin{equation}
h_{il}(t)=\eta_dh_{ild}(t)
\label{hil}
\end{equation}

The model in (\ref{Yc}) is a hybrid piece affine model with three modes. The injection mode denotes that the vehicle is being injected. The generation mode represents that the vehicle is available for power generation. The driving mode denotes that the vehicle is driving. The three modes will not happen simultaneously. $D_{iyl}(t)+d_{il}(t)$ is the hydrogen injected into the vehicle $l_i$ at time slot $t$. Fuel cell vehicle $l_i$ obtains hydrogen $D_{iyl}(t)$ from MG $i$ and purchases hydrogen $d_{il}(t)$ from the hydrogen station of the hydrogen-producing company. $Y_{ifl}(t)$ is the hydrogen consumed for generation by vehicle $l_i$ at time slot $t$. 
The power generated by fuel cell vehicle $l_i$ is denoted as $\eta_fhY_{ifl}(t)$, where $\eta_f$ is the conversion efficiency of the fuel cell from hydrogen to electricity. $h_{il}(t)$ is the hydrogen used for transportation by vehicle $l_i$ at time slot $t$, $h_{ild}(t)$ is the travel distance, and $\eta_d$ is the hydrogen by each vehicle consumes per kilometer. For fuel cell vehicle $l_i$, there are maximum hydrogen storage $Y_{il,max}$, hydrogen injected $D_{iyl,max}$ and $d_{il,max}$, hydrogen consumed for generation $Y_{ifl,max}$, and hydrogen used for transportation $h_{il,max}$ during one time slot:
\begin{equation}
0\leq Y_{il}(t)\leq Y_{il,max}
\label{Yil}
\end{equation}
\begin{equation}
0\leq D_{iyl}(t)\leq D_{iyl,max}, 
0\leq d_{il}(t)\leq d_{il,max}
\label{Yi1}
\end{equation}
\begin{equation}
0\leq Y_{ifl}(t)\leq Y_{ifl,max}
\label{Yi2}
\end{equation}
\begin{equation}
0\leq h_{il}(t)\leq h_{il,max}
\label{Yi3}
\end{equation}

\subsection{Cost Function}
 The cost function of MG $i$ consists of the payment and revenue, which is denoted as
\begin{equation}
\begin{aligned}
C_{i}(t)&=C_{ihy}(t)+C_{ip}(t)+C_{ig}(t)+C_{iX}(t)-R_{iS}(t)-R_{ip}(t)
\end{aligned}
\label{micro1}
\end{equation}

\begin{equation}
\begin{aligned}
C_{ihy}(t)&=p_{y}(t)d_{i}(t),
C_{ip}(t)=E_{i}(t)p_{e}(t) \\
C_{ig}(t)&=(P_{i}^{CHP}(t)+H_{i}^{CHP}(t)+H_{i}^b(t))p_{g}(t) \\
C_{iX}(t)&=\beta_i(t)X_{i}(t),
R_{iS}(t)=\alpha_i(t)S_{i}(t),
R_{ip}(t)=E_{io}(t)p_{eo}(t) 
\end{aligned}
\end{equation}
where $C_{ihy}(t)$ is the hydrogen cost of purchasing hydrogen from the hydrogen-producing company by all vehicles of MG $i$ at time slot $t$. $C_{ip}(t)$ and $C_{ig}(t)$ are the costs of purchasing electricity and gas from the electricity and gas utility company at time slot $t$. $C_{iX}(t)$ and $R_{iS}(t)$ are the cost of purchasing electricity from other MGs and the revenue of selling electricity to other MGs in energy trading at time slot $t$. $R_{ip}(t)$ is the revenue from selling electricity to the electricity utility company at time slot $t$. $p_{y}(t)$ is the hydrogen price of the hydrogen-producing company. $d_{i}(t)$ is the amount of hydrogen purchased from the hydrogen-producing company by all vehicles of MG $i$ at time slot $t$. $E_{i}(t)$ is the amount of electricity purchased from the electricity utility company by MG $i$ at time slot $t$. When MG $i$ purchases electricity from other MGs, $\beta_i(t)$ is the purchase price and $X_{i}(t)$ is the amount of electricity at time slot $t$. When MG $i$ sells electricity to other MGs, $\alpha_i(t)$ is the selling price of MG $i$ and $S_{i}(t)$ is the amount of electricity at time slot $t$. $E_{io}(t)$ and $p_{eo}(t)$ are the amount and price of electricity sold to the electricity utility company by MG $i$ at time slot $t$. 

Note that the electricity demand $L_{ie}(t)$, hydrogen demand $L_{iy}(t)$, and heat demand $L_{ih}(t)$ of MG $i$ should be satisfied when they arrive, i.e.
\begin{equation}
\begin{aligned}
L_{ie}(t)&=E_{i}(t)+N_{i}(t)+X_{i}(t)-S_{i}(t)
-C_{ie}(t)+D_{ie}(t)\\&+\eta_fhY_{if}(t)-\frac{hC_{iy}(t)}{\eta_e}-c_1C_{iy}(t)+\eta_{pg}P_{i}^{CHP}(t)-E_{io}(t)\\
L_{iy}(t)&=h_{i}(t)\\
L_{ih}(t)&=\eta_{hg}H_{i}^{CHP}(t)+\eta_{bg}H_{i}^b-C_{ih}(t)+D_{ih}(t)
\end{aligned}
\label{lg}
\end{equation}

\section{Solution Methodology}
\subsection{Optimization Method}
 The strategy set of MG $i$ is $\boldsymbol{M}_{i}(t)$=\{$C_{ie}(t)$, $D_{ie}(t)$, $C_{iy}(t)$, $D_{iy}(t)$, $C_{ih}(t)$, $D_{ih}(t)$, $D_{iyl}(t)$, $d_{il}(t)$, $Y_{ifl}(t)$, $h_{il}(t)$, $E_{i}(t)$,   $P^{CHP}_{i}(t)$, $H^{CHP}_{i}(t)$, $H_{i}^{b}(t)$, $X_{i}(t)$, $S_{i}(t)$, $E_{io}(t)$\}. According to the system model, the optimization problem of MG $i$ is to find a control policy that schedules the electricity, hydrogen, and heat at each time slot to minimize the time average energy cost, which can be denoted as a stochastic network optimization problem: 
\begin{equation}
\begin{aligned}
&\min_{\boldsymbol{M}_{i}(t)}  \lim_{T \rightarrow \infty}\frac{1}{T} \sum_{t=1}^{T} \mathbb{E}\{C_{i}(t)\}
\end{aligned}
\label{eq30}
\end{equation}
subject to (\ref{A1}) - (\ref{Yi3}) and (\ref{lg}).

{\color{black}{The Lyapunov optimization gives simple online solutions based on the current information of the system state as opposed to traditional approaches like Markov decision processes and dynamic programming, which have very high computational complexity and require a priori information of all the random processes in the system. The performance of the Lyapunov optimization algorithm can be close to the optimal value arbitrarily \cite{Lakshminarayana2014Cooperation}. The underlying assumption about the availability of future information renders offline approaches ill-suited for energy storage system applications with high uncertainty, whereas dynamic programming solutions are impractical for multiple networked energy storage systems \cite{Sarthak2018Optimal}.}}
The time average expected values under any feasible control policy of the original problem are denoted as follows:
\begin{equation}
\begin{split}
\overline{C_{ie}}&=\lim_{T\rightarrow\infty}\frac{1}{T}\sum_{t=1}^{T}\mathbb{E}\{C_{ie}(t)\},
\overline{D_{ie}}=\lim_{T\rightarrow\infty}\frac{1}{T}\sum_{t=1}^{T}\mathbb{E}\{D_{ie}(t)\}\\
\overline{C_{iy}}&=\lim_{T\rightarrow\infty}\frac{1}{T}\sum_{t=1}^{T}\mathbb{E}\{C_{iy}(t)\} ,
\overline{D_{iy}}=\lim_{T\rightarrow\infty}\frac{1}{T}\sum_{t=1}^{T}\mathbb{E}\{D_{iy}(t)\}\\
\overline{C_{ih}}&=\lim_{T\rightarrow\infty}\frac{1}{T}\sum_{t=1}^{T}\mathbb{E}\{C_{ih}(t)\},
\overline{D_{ih}} =\lim_{T\rightarrow\infty}\frac{1}{T}\sum_{t=1}^{T}\mathbb{E}\{D_{ih}(t)\}\\
\overline{D_{iyl}}&+\overline{d_{il}}=\lim_{T\rightarrow\infty}\frac{1}{T}\sum_{t=1}^{T}\mathbb{E}\{D_{iyl}(t)+d_{il}(t)\}\\
 \overline{Y_{ifl}}&+\overline{h_{il}}=\lim_{T\rightarrow\infty}\frac{1}{T}\sum_{t=1}^{T}\mathbb{E}\{Y_{ifl}(t)+h_{il}(t)\}
\end{split}
\label{sto1}
\end{equation}

The above stochastic network optimization problem (\ref{eq30}) cannot be solved directly owing to the capacity constraints of
battery, hydrogen storage, and water tank (\ref{Bm}) - (\ref{Wm}) of MG $i$ and hydrogen storage (\ref{Yil}) of fuel cell vehicle $l_i$. To be specific, stochastic network optimization can ensure that the average energy consumption equals the average energy generation for a long period, but cannot provide a hard constraint on the difference between consumption and generation at any time slot. In order to solve this issue, the problem is relaxed,
which is stated as follows: 
The optimization problem (\ref{eq30}) is subject to
\begin{equation}
\begin{split}
\overline{C_{ie}} &= \overline{D_{ie}}\\
\overline{C_{iy}} &= \overline{D_{iy}}\\
\overline{C_{ih}} &= \overline{D_{ih}}\\
\overline{D_{iyl}}+\overline{d_{il}} &= \overline{Y_{ifl}}+\overline{h_{il}}\\
\end{split}
\label{cd}
\end{equation}
and (\ref{Cem}) - (\ref{Chm}), (\ref{DdYh}), (\ref{hil}), (\ref{Yi1}) - (\ref{Yi3}), (\ref{lg}).

$C_{i}^{opt}$ is denoted as the optimal solution of the cost function for the original problem and $C_{ir}^{opt}$ is denoted as the optimal solution of the cost function for the relaxed problem. Any feasible solution to the original problem is also a feasible solution to the relaxed problem, that is, the relaxed problem is less constrained than the original problem. Therefore, $C_{ir}^{opt} \leq C_{i}^{opt}$. 

The optimal solution to the relaxed problem can be got by the stationary and randomized policy $\Pi$, stated as follows:
\begin{equation}
\begin{aligned}
\mathbb{E}\{C_{i}^{\Pi}(t)\}=C_{ir}^{opt}
\end{aligned}
\end{equation}
subject to: 
\begin{equation}
\begin{split}
C^{\Pi}_{ie}(t)& = D^{\Pi}_{ie}(t)\\
C^{\Pi}_{iy}(t) &= D^{\Pi}_{iy}(t)\\
C^{\Pi}_{ih}(t) &= D^{\Pi}_{ih}(t)\\
D^{\Pi}_{iyl}(t)+d^{\Pi}_{il}(t) &= Y^{\Pi}_{ifl}(t)+h^{\Pi}_{il}(t)\\
\end{split}
\label{cd2}
\end{equation} and  (\ref{Cem}) - (\ref{Chm}), (\ref{DdYh}), (\ref{hil}), (\ref{Yi1}) - (\ref{Yi3}), (\ref{lg}).

The existence of the stationary and randomized policy can be proved by the Caratheodory theory \cite{georgiadis2006resource}. 
 Obviously, only if the solutions to the relaxed problem can meet constraints (\ref{Bm}) - (\ref{Wm}) and (\ref{Yil}), they are also feasible to the original problem. To reach this objective, the constants $\theta_{i}$, $\xi_{i}$, $\varepsilon_{i}$ and $\gamma_{il}$ are defined. These constants are adjusted appropriately to make the solutions to the relaxed problem also be feasible to the original problem. To start, the virtual queues $A_{i}(t)$, $F_{i}(t)$, $Z_{i}(t)$, and $I_{il}(t)$ for battery, hydrogen storage, water tank of MG $i$, and hydrogen storage of fuel cell vehicle $l_i$ are defined as follows, respectively:
\begin{equation}
\begin{aligned}
&A_{i}(t)=B_{i}(t)-\theta_{i} ,
F_{i}(t)=Y_{i}(t)-\xi_{i} \\
&Z_{i}(t)=W_{i}(t)- \varepsilon_{i},
I_{il}(t)=Y_{il}(t)- \gamma_{il}
\end{aligned}
\label{vir}
\end{equation}
where $\theta_{i}$, $\xi_{i}$ and $\varepsilon_{i}$ and $\gamma_{il}$ are perturbations that are used to guarantee the bounds of $B_{i}(t)$, $Y_{i}(t)$, $W_{i}(t)$, and $Y_{il}(t)$. 

The Lyapunov function is defined as $Q_{i}(t)=\frac{1}{2}A_{i}(t)^{2}+\frac{1}{2}F_{i}(t)^{2}+\frac{1}{2}Z_{i}(t)^{2}+\frac{1}{2}\sum^{L_i}_{l=1}I_{il}(t)^{2}$. The conditional Lyapunov drift, which represents the change in the Lyapunov function, is defined as
\begin{equation}
\Delta_{i}(t)=\mathbb{E}\{Q_{i}(t+1)-Q_{i}(t)|B_{i}(t),Y_{i}(t),W_{i}(t), Y_{il}(t)\}
\end{equation}
where the expectation is related to the random processes of the system, given the values $B_{i}(t)$, $Y_{i}(t)$, $W_{i}(t)$, and $Y_{il}(t)$.  According to the equation for the virtual queue (\ref{vir}) associated with the evolution of battery, hydrogen storage, and water tank of MG $i$ in  (\ref{A1}) - (\ref{A3}), and the hydrogen storage of the fuel cell vehicle in (\ref{Yc}), the Lyapunov drift is bounded as
\begin{equation}
\begin{aligned}
\Delta_{i}(t)&=\mathbb{E}\{Q_{i}(t+1)-Q_{i}(t)|B_{i}(t),Y_{i}(t),W_{i}(t), Y_{il}(t)\} 
\\&\leq G_{i}+\mathbb{E}\{A_{i}(t)(C_{ie}(t)-D_{ie}(t))+F_{i}(t)(C_{iy}(t)-D_{iy}(t))\\&+Z_{i}(t)(C_{ih}(t)-D_{ih}(t))+\sum^{L_i}_{l=1}[I_{il}(t)(D_{iyl}(t)+d_{il}(t)-Y_{ifl}(t)-h_{il}(t))]\}
\end{aligned}
\label{rightmin}
\end{equation}
where $G_{i}$ is constant and $G_{i}=\frac{1}{2}\{\max(C_{ie,max}^2,D_{ie,max}^2)+\max(C_{iy,max}^2,D_{iy,max}^2)+\max(C_{ih,max}^2,D_{ih,max}^2)+\sum^{L_i}_{l=1}[\max((D_{iyl,max}+d_{il,max})^2,(Y_{ifl,max}+h_{il,max})^2)]\}$.
The proof of this step is given in Appendix A.

In order to make these queues stable, MG $i$ needs to minimize drift $\Delta_{i}(t)$. In addition, MG $i$ intends to minimize the energy cost. Hence, $V_{i}$ is used to represent the tradeoff between the two objectives. Then, the drift-plus-penalty function is denoted as
\begin{equation}
\begin{aligned}
\Delta_{i}(t)+V_{i}\mathbb{E}\{C_{i}(t)\} \leq & G_{i}+\mathbb{E}\{A_{i}(t)(C_{ie}(t)-D_{ie}(t))+F_{i}(t)(C_{iy}(t)-D_{iy}(t))\\&+Z_{i}(t)(C_{ih}(t)-D_{ih}(t))+\sum^{L_i}_{l=1}[I_{il}(t)(D_{iyl}(t)+d_{il}(t)\\&-Y_{ifl}(t)-h_{il}(t))]\}+V_{i}\mathbb{E}\{C_{i}(t)\}
\\=&G_{i}+\mathbb{E}\{A_{i}(t)(C_{ie}(t)-D_{ie}(t))
+F_{i}(t)(C_{iy}(t)-D_{iy}(t))\\&+Z_{i}(t)(C_{ih}(t)-D_{ih}(t))+\sum^{L_i}_{l=1}[I_{il}(t)(D_{iyl}(t)+d_{il}(t)\\&-Y_{ifl}(t)-h_{il}(t))]\
+V_{i}(
p_{y}(t)d_{i}(t)
+E_{i}(t)p_{e}(t)\\&+(P_{i}^{CHP}(t)+H_{i}^{CHP}(t)+H_{i}^b(t))p_{g}(t)
+\beta_i(t)X_{i}(t) \\&-\alpha_i(t)S_{i}(t)-E_{io}(t)p_{eo}(t))\}
\end{aligned}
\label{p1}
\end{equation}

The relaxed problem can be viewed as minimizing the cost of the MG while maintaining the stability of virtual queues. The drift-plus-penalty term consists of two terms: the Lyapunov drift term $\Delta_{i}(t)$ and the modified cost term $V_{i}\mathbb{E}\{C_{i}(t)\}$. 
A larger value of $V_{i}$ means that minimizing the energy cost has greater priority than minimizing the drift, and vice versa.  The objective of Lyapunov optimization is to minimize the right hand of (\ref{p1}), i.e. 
\begin{equation}
\begin{aligned}
&\min_{\boldsymbol{M}_{i}(t)} A_{i}(t)(C_{ie}(t)-D_{ie}(t))+F_{i}(t)(C_{iy}(t)-D_{iy}(t))\\&+Z_{i}(t)(C_{ih}(t)-D_{ih}(t))+\sum^{L_i}_{l=1}[I_{il}(t)(D_{iyl}(t)+d_{il}(t)\\&-Y_{ifl}(t)-h_{il}(t))]
+V_{i}(
p_{y}(t)d_{i}(t)
+E_{i}(t)p_{e}(t)\\&+ (P_{i}^{CHP}(t)+H_{i}^{CHP}(t)+H_{i}^b(t))p_{g}(t)+\beta_i(t) X_{i}(t)\\&-\alpha_i(t) S_{i}(t)-E_{io}(t)p_{eo}(t))
\end{aligned}
\label{P2}
\end{equation}
subject to constraints (\ref{Cem}) - (\ref{Chm}), (\ref{DdYh}), (\ref{hil}), (\ref{Yi1}) - (\ref{Yi3}), (\ref{lg}).

In the following section, the price and amount of energy in energy trading among multiple MGs are determined, and the optimal strategy of problem (\ref{P2}) is obtained by solving the linear programming problem.

\subsection{Double-Auction Mechanism}
Optimization problem (\ref{P2}) has two variables of price. Owing to the decentralized structure of energy trading, the selling price and purchase price can be determined by the external auctioneer according to a double-auction mechanism.
First, the selling price and purchase price of each MG submitted in energy trading among multiple MGs are investigated.

\textbf{Lemma 1.} 
MG $i$ decides the selling price $\widetilde{\alpha}_{i}(t)$ and purchase price $\widetilde{\beta}_{i}(t)$ based on the cost-minimization problem:
\begin{equation}
\widetilde{\alpha}_{i}(t)=\max[\frac{-A_{i}(t)}{V_{i}},\frac{-F_{i}(t)}{(\frac{h}{\eta_{e}}+c_1)V_{i}}, p_{eo}(t)]
\label{e11}
\end{equation}
\begin{equation}
\widetilde{\beta}_{i}(t)=\min[\frac{\max(-A_{i}(t),0)}{V_{i}},\frac{p_{g}}{\eta_{pg}},p_{e}(t)]
\label{e12}
\end{equation}
where $p_{eo}(t)$ is the price of energy sold to the electricity utility company by MGs, and $p_{eo}(t)<p_{e}(t)$ .

The proof of this step is presented in Appendix B.



After determining $\widetilde{\alpha}_{i}(t)$ and $\widetilde{\beta}_{i}(t)$, the amount of electricity $\widetilde{S}_{i}(t)$ and $\widetilde{X}_{i}(t)$ that MG $i$ will sell and purchase in energy trading are determined by solving (\ref{P2}). MGs are willing to sell their energy when their energy storages have enough energy. Moreover, they are willing to get energy when the cost of purchasing energy is lower than that of generating energy by themselves (such as generating electricity by CHP system and using hydrogen). The maximum amount of electricity that MG $i$ can sell $S_{i,max}(t)$ and purchase $X_{i,max}(t)$ at time slot $t$ are:
\begin{equation}
S_{i,max}(t)=N_{i}(t)-L_{ie}(t)
\end{equation}
\begin{equation}
X_{i,max}(t)=L_{ie}(t)-N_{i}(t)
\end{equation}

A double-auction mechanism is designed to encourage MGs to actively trade energy and ensure the benefits of MGs.
The double-auction mechanism has two steps:
\begin{itemize}
\item MGs submit the selling price, purchase price, and the corresponding amount of energy to the external auctioneer.
\item The external auctioneer decides the accepted selling price and purchase price by trading rules, and allocates energy to MGs to minimize the transmission loss.
\end{itemize}


The mechanism of the threshold price double auction \cite{Kant2005Double} is shown in this section. First, the external auctioneer collects and sorts all received purchase prices in descending order and selling prices in ascending order:
$\overline{\beta}_{1}(t)\geq \overline{\beta}_{2}(t)\geq\cdots\geq \overline{\beta}_{i}(t) \geq r>\overline{\beta}_{i+1}(t)  \geq\cdots \geq\overline{\beta}_{n}(t)$ and $\overline{\alpha}_{1}(t)\leq  \overline{\alpha}_{2}(t)\leq \cdots\leq \overline{\alpha}_{j}(t)\leq r< \overline{\alpha}_{j+1}(t)\leq \cdots \leq\overline{\alpha}_{n}(t)$.  If $i=j$, the external auctioneer notifies MG $l$, $l=1,2, \cdots , i$, that they can trade with. The accepted selling price and purchase price are the same, i.e., $\alpha(t)=\beta(t)=r$. If $i>j$, the external auctioneer notifies MG $l$, $l=1,2, \cdots , j$, that they can trade with. The accepted selling price and purchase price are $\alpha(t)=r$ and $\beta(t)=\overline{\beta}_{j+1}(t)$, respectively. If $i<j$, the external auctioneer notifies MG $l$, $l=1,2, \cdots , i$, that they can trade with. The accepted selling price and purchase price are $\alpha(t)=\overline{\alpha}_{i+1}(t)$ and $\beta(t)=r$, respectively. 

The accepted purchase price and selling price for MG $i$ can be derived as
\begin{equation}
\hat{\beta}_{i}(t)=\left\{\begin{array}{cc}
\beta(t) & \mbox{if MG $i$ purchases electricity}\\
0 & \mbox{otherwise} \end{array}\right.
\end{equation}
and
\begin{equation}
\hat{\alpha}_{i}(t)=\left\{\begin{array}{cc}
\alpha(t) & \mbox{if MG $i$ sells electricity}\\
0 & \mbox{otherwise} \end{array}\right.
\end{equation}

After determining the market clearing prices, the external auctioneer needs to match energy sellers and buyers to reduce energy losses.

\begin{equation}
Loss(t) = \sum_{i=1}^{k}\sum_{j=1}^{k} I_{ij}T_{ij}(t)
\end{equation}
where $T_{ij}(t)$ is the amount of energy transmitted from MG $i$ to MG $j$ at time $t$. $I_{ij}$ is the energy loss coefficient, which is related to the transmission distance. 

The external auctioneer aims to minimize the energy losses during transmission:
\begin{equation}
\min_{T_{ij}, \forall i, j \in [1, k]} \quad Loss(t)
\label{chap3equ:minLoss}
\end{equation}
subject to:
\begin{equation}
\sum_{j=1}^{k} T_{ij} \leq \overline{S}_{i}(t)
\label{chap3equ:sell}
\end{equation}
\begin{equation}
\sum_{i=1}^{k} (1-I_{ij})T_{ij} \geq \overline{X}_{j}(t)
\label{chap3equ:buy}
\end{equation}

After determining $\alpha_{i}(t)$ and $\beta_{i}(t)$, the actual amount of electricity that MG $i$ sells $S^{*}_{i}(t)$ or purchases $X^{*}_{i}(t)$ can be determined to minimize the energy losses during transmission by linear programming. 
The performance of the proposed trading mechanism is as follows:





\textbf{Lemma 2.} Using the mechanism presented above, all MGs will submit the selling prices and purchase prices truthfully; otherwise, they will get lower benefits owing to deviating from the true value of the selling prices and purchase prices in (\ref{e11}) and (\ref{e12}).

The proof of this step is shown in Appendix C.

\subsection{Algorithm Design and Performance Analysis}
After obtaining $X^{*}_{i}(t)$ and $S^{*}_{i}(t)$ by (\ref{chap3equ:minLoss})-(\ref{chap3equ:buy}) and the double-auction mechanism, an optimal strategy set of MG $i$ can be acquired by solving the linear programming problem (\ref{P2}):
$\boldsymbol{M}_{i}^{*}(t)$=\{$C_{ie}^{*}(t)$, $D_{ie}^{*}(t)$, $C_{iy}^{*}(t)$, $D_{iy}^{*}(t)$, $C_{ih}^{*}(t)$, $D_{ih}^{*}(t)$, $D_{iyl}^{*}(t)$, $d_{il}^{*}(t)$, $Y_{ifl}^{*}(t)$, $h_{il}^{*}(t)$, $E_{i}^{*}(t)$, $P^{CHP,*}_{i}(t)$, $H^{CHP,*}_{i}(t)$, $H_{i}^{b,*}(t)$, $X^{*}_{i}(t)$, $S^{*}_{i}(t)$, $E_{io}^{*}(t)$
\} to minimize the drift-plus-penalty.
The implementation process of the algorithm is shown in Algorithm 1.

\begin{algorithm}[h]
\caption{Joint Energy Scheduling and Trading Algorithm}
\begin{algorithmic}[1]
\State Set $t=0$.
\State Set the initial values $B_{i}(t)$, $Y_{i}(t)$, $W_{i}(t)$, and $Y_{il}(t)$.
\For{each MG $i$}
\State Calculate $\overline{\alpha}_{i}(t)$ and $\overline{\beta}_{i}(t)$ by (\ref{e11}) and (\ref{e12}), calculate $\overline{X}_{i}(t)$ and $\overline{S}_{i}(t)$ by (\ref{P2}), and then submit them to the external auctioneer.
\State Calculate $\alpha_{i}(t)$, $\beta_{i}(t)$, $X_{i}(t)$, and $S_{i}(t)$ by the double-auction mechanism.
\State Calculate $\boldsymbol{M}_i(t)$ using (\ref{P2}).
\EndFor
\State Update $B_{i}(t+1)$, $Y_{i}(t+1)$, $W_{i}(t+1)$ by (\ref{A1}) - (\ref{A3}), and $Y_{il}(t+1)$ by (\ref{Yc}).
\label{code:recentEnd}
\end{algorithmic}
\end{algorithm}

In the aforementioned design, the capacity constraints of battery, hydrogen storage, and water tank of MG $i$ and hydrogen storage of fuel cell vehicle $l_i$ are not considered. In fact, the capacity constraints should be considered as follows:

\textbf{Lemma 3.} If $\theta_{i}$, $\xi_{i}$, $\varepsilon_{i}$, $\gamma_{il}$ and $V_{i}$ satisfy the following conditions:
\begin{equation}
\theta_{i}=V_{i}p_{e,max}+D_{ie,max}
\label{A2}
\end{equation}
\begin{equation}
\xi_{i}=V_{i}p_{y,max}+D_{iy,max}
\label{Y2}
\end{equation}
\begin{equation}
\varepsilon_{i}=\frac{V_{i}p_{g,max}}{\eta_{bg}}+D_{ih,max}
\label{A4}
\end{equation}
\begin{equation}
\gamma_{il}=V_{i}p_{y,max}+Y_{ifl,max}+h_{il,max}
\label{A5}
\end{equation}
\begin{equation}
\begin{aligned}
V_{i,max}= & \min \{\frac{B_{i,max}-C_{ie,max}-D_{ie,max}}{p_{e,max}}, \\&\frac{Y_{i,max}-C_{iy,max}-D_{iy,max}}{p_{y,max}}, \\& \frac{\eta_{bg}(W_{i,max}-C_{ih,max}-D_{ih,max})}{p_{g,max}}, \\&\frac{Y_{il,max}-D_{iyl,max}-d_{il,max}-Y_{ifl,max}-h_{il,max}}{p_{y,max}}\}\\
\end{aligned}
\label{P3}
\end{equation}
where $0\leq V_{i} \leq V_{i,max}$, the capacity constraints of battery, hydrogen storage, and water tank of MG $i$ and hydrogen storage of fuel cell vehicle $l_i$ are always satisfied. 

The proof of this step is presented in Appendix D. 

According to Lemma 3, the algorithm satisfies the capacity constraints in (\ref{Bm}) - (\ref{Wm}) and (\ref{Yil}). Hence, the algorithm is feasible for the original problem.  Then, the result of the performance of the algorithm based on Lyapunov optimization is provided.

\textbf{Theorem 1.} According to the algorithm in the previous section, the expected time average energy cost has a bound:
\begin{equation}
\lim_{T \rightarrow \infty} \frac{1}{T} \sum_{t=1}^{T} \mathbb{E}\{ C_{i}(t) \} \leq C_{i}^{opt} + \frac{G_{i}}{V_{i}}
\end{equation}

The proof of this step is given in Appendix E. 
{\color{black}{In a sense, Theorem 1 shows the gap between the performance of the proposed algorithm, independent of random distribution factors, and an optimization algorithm with accurate random information is described. According to (\ref{P3}) and Theorem 1, battery, hydrogen storage, and water tank capacity of MG $i$ and hydrogen storage capacity of fuel cell vehicle $l_i$ increase, the performance of the proposed algorithm can be made arbitrarily close to the optimal performance of the optimization algorithm with accurate random information.}}

\section{Numerical Results}

In this section, the numerical results based on real data are presented to examine the proposed algorithm in the previous sections.
\subsection{Experimental Setup}

\begin{figure*}
\centering
\begin{minipage}{0.33\linewidth}
  \centerline{\includegraphics[height=30mm,width=40mm]{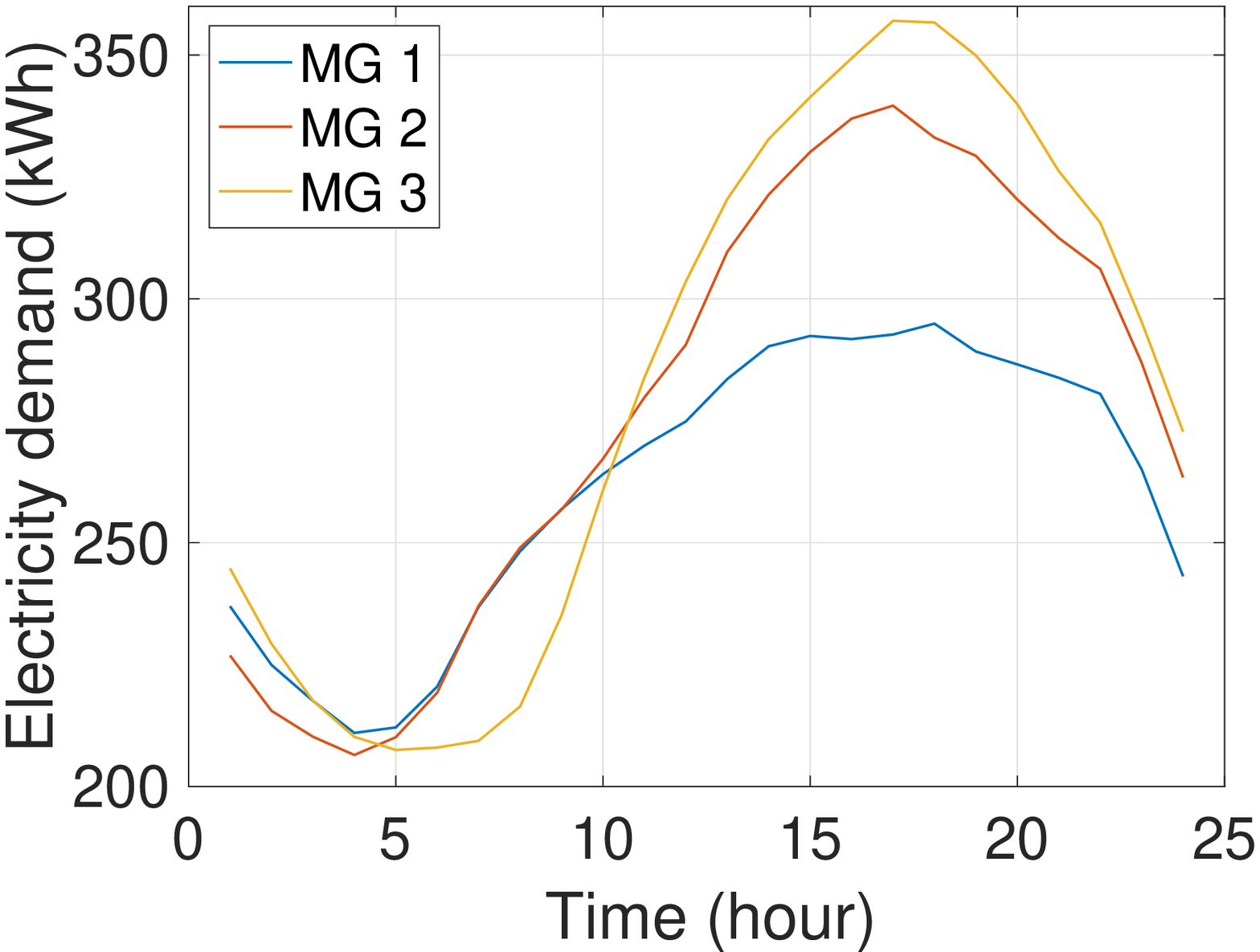}}
  \centerline{\scriptsize{(a) Electricity demand}}
\end{minipage}
\hspace{-5pt}
\begin{minipage}{0.33\linewidth}
 \centerline{\includegraphics[height=30mm,width=40mm]{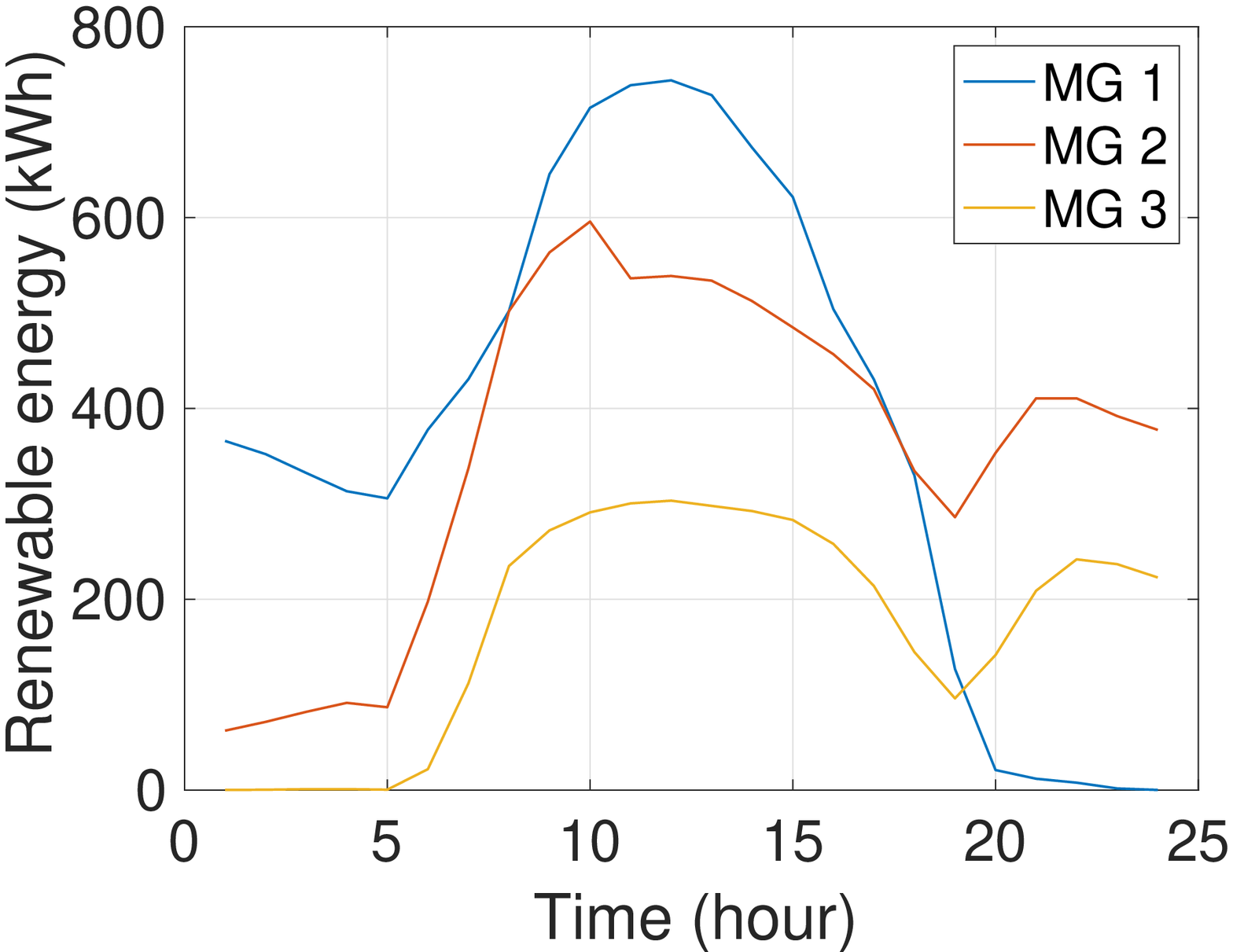}}
  \centerline{\scriptsize{(b) Renewable energy }}
\end{minipage}
\hspace{-5pt}
\begin{minipage}{0.33\linewidth}
  \centerline{\includegraphics[height=30mm,width=40mm]{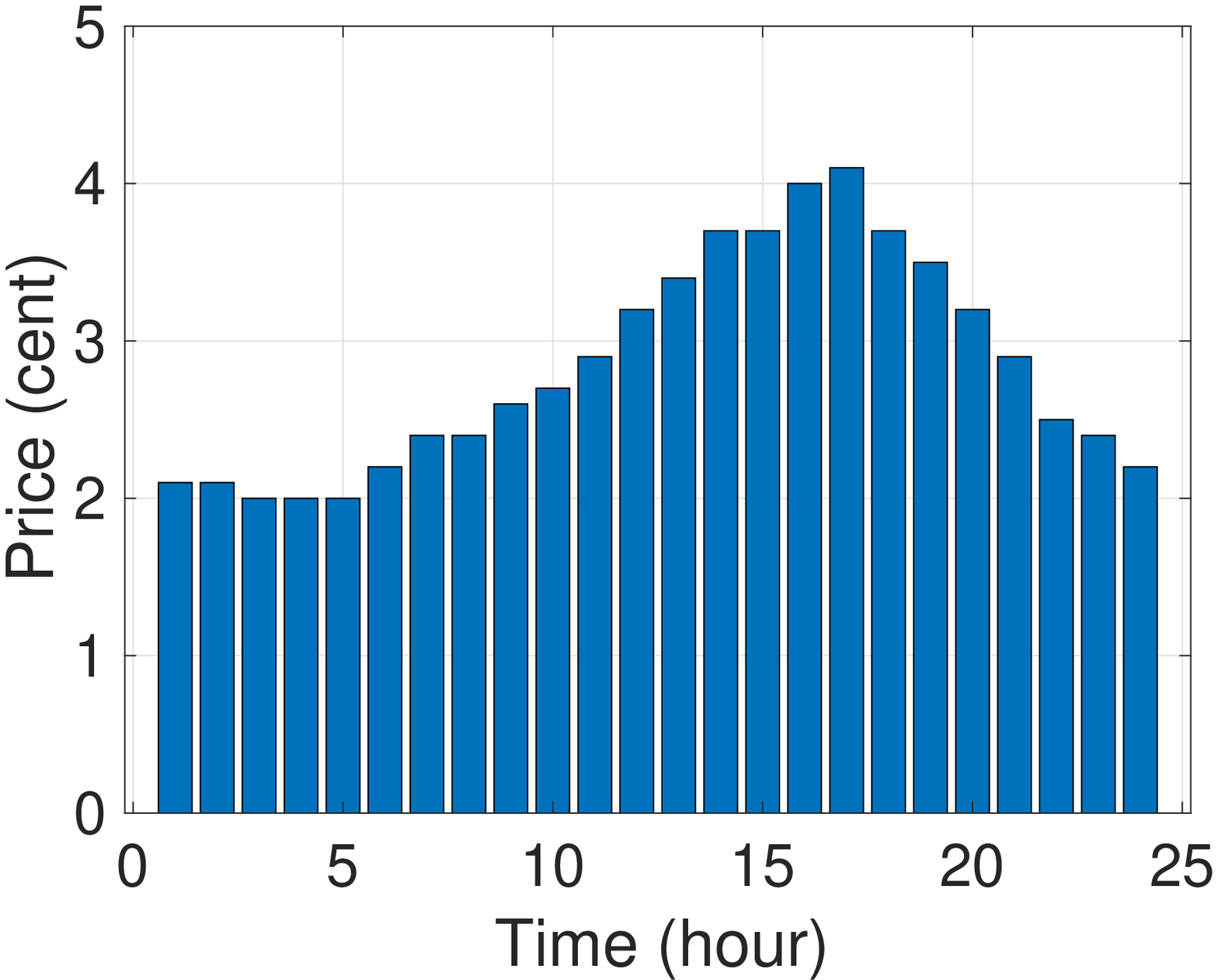}}
  \centerline{\scriptsize{(c) Price of the electricity utility company}}
\end{minipage}
\caption{Data from website.}
\label{fig2}
\end{figure*}

A network of three MGs is considered. Each MG includes renewable energy resources, CHP system, fuel cell vehicles, battery, hydrogen storage, boiler, and water tank. 
Wind-driven turbines and photovoltaic systems are renewable energy generators with maximum outputs of 750 kW for MGs $1$ and $2$, and 450 kW for MG $3$. 
For each MG's electricity load, the hourly load data provided by the PJM hourly load \cite{pjm} is shown in Fig. \ref{fig2}(a). 
For renewable energy generation, the hourly generation data provided by Renewables.ninja \cite{Institute} is shown in Fig. \ref{fig2}(b). 
For the price of the electricity utility company, the hourly energy price provided by the Power Smart Pricing administered for Ameren Illinois data \cite{Illinois} is shown in Fig. \ref{fig2}(c). 

The maximum electricity consumption of the electrolysis system is 100 kW. The total amount of hydrogen consumed by each MG's all vehicles, which are used for driving, takes value from [25, 35] m$^3$ at random during a time slot. When hydrogen in the hydrogen storage of the MG cannot supply the vehicles, the vehicles will purchase hydrogen from the hydrogen station of the hydrogen-producing company. In this paper, fuel cell vehicles refer to buses. Each MG has 10 fuel cell buses. The hydrogen by each bus consumes is approximately 0.5 m$^3$/km, and the maximum generation of the bus is 45 kW. The buses depart every 20 minutes from 6:00 am to 10:00 pm. At 6:00 am, there are 5 buses at the starting point and the bus terminal of each MG, respectively. The distance from the starting point to the bus terminal is about 10km, and it takes 40-60 minutes to complete the journey.

The heat and electricity generation of CHP system satisfies $H_{i}^{CHP}(t)=P_{i}^{CHP}(t)$. 
The parameters of efficiency are $\eta_{pg}=70\%$, $\eta_{hg}=70\%$, $\eta_{bg}=80\%$, $\eta_{e}=85\%$, and $\eta_{f}=50\%$, respectively. Other parameters are summarized as follows: 
$p_{y}(t)=10$ cents/m$^3$, $p_{g}(t)=15$ cents/m$^3$, $B_{i,max}=300$kWh, $C_{ie,max}=D_{ie,max}=75$kWh, $W_{i,max}=900$kWh, $C_{ih,max}=D_{ih,max}=225$kWh.
$Y_{i,max}=300$m$^3$, $C_{iy,max}=D_{iy,max}=75$m$^3$. 

\subsection{Results}
\begin{figure}
\centering
  \centerline{\includegraphics[height=42mm,width=56mm]{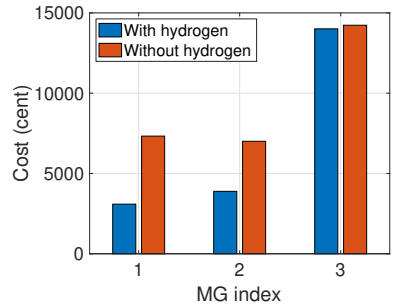}}
  \caption{Comparisons of all MGs' total costs with and without hydrogen storage.}
  \label{fig3}
\end{figure}

Fuel cell vehicles and hydrogen storages play important roles in relieving storage stress of the battery and further using excess renewable energy. Fig. \ref{fig3} shows that the costs of the MGs are lower than those without hydrogen storage. {
{The existence of hydrogen storage obviously reduces the costs of MGs 1 and 2. The cost of MG 3, however, is slightly reduced. The reason is because MGs 1 and 2 electrolyze water to supply hydrogen for fuel cell vehicles instead of selling electricity to the electricity company at a low price. Therefore, the costs of MGs 1 and 2 are obviously reduced. Because the renewable energy of MG 3 is not enough, MG 3 needs to purchase energy. 
In Fig. \ref{fig6}(c), MG 3 with hydrogen storage electrolyzes water to supply a little hydrogen for fuel cell vehicles, and MG 3 without hydrogen storage charges the battery. Both of them purchase much hydrogen from the hydrogen-producing company. Therefore, the cost of MG 3 with hydrogen is slightly reduced in Fig. \ref{fig3}. }}

\begin{figure*}
\centering
\begin{minipage}{0.33\linewidth}
  \centerline{\includegraphics[height=3cm,width=4cm]{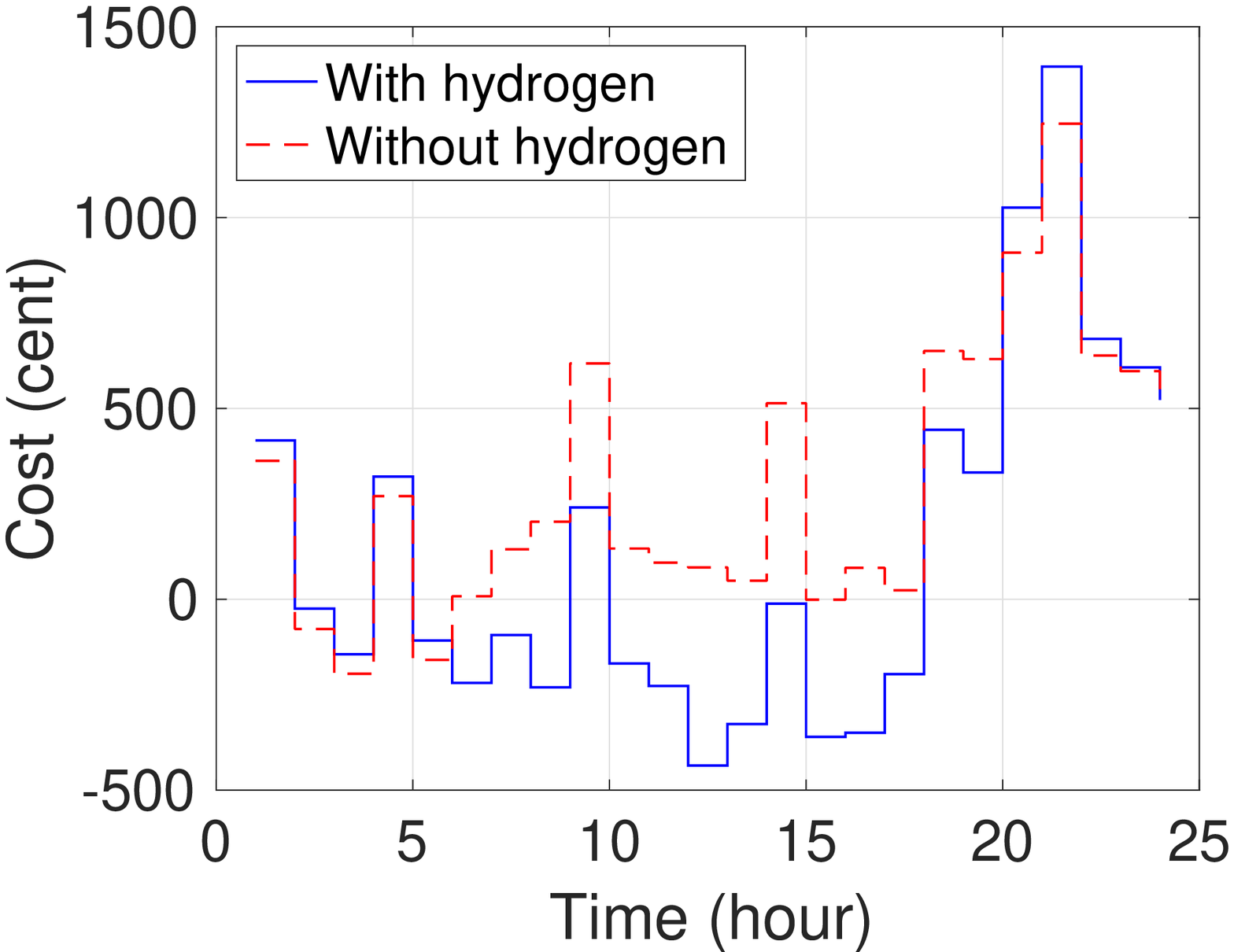}}
  \centerline{\scriptsize{(a) MG 1}}
\end{minipage}
\hspace{-5pt}
\begin{minipage}{0.33\linewidth}
  \centerline{\includegraphics[height=3cm,width=4cm]{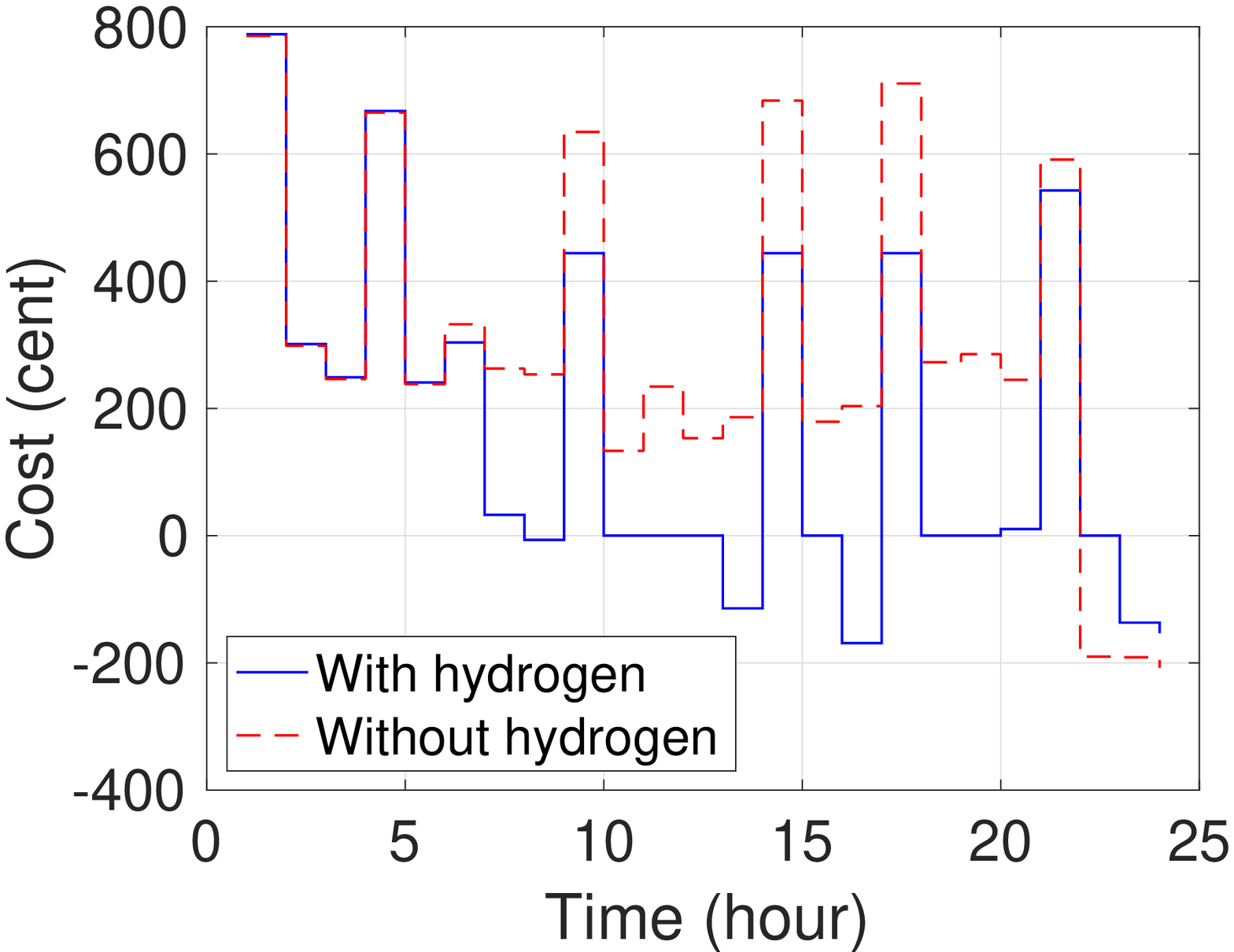}}
  \centerline{\scriptsize{(b) MG 2}}
\end{minipage}
\hspace{-5pt}
\begin{minipage}{0.33\linewidth}
  \centerline{\includegraphics[height=3cm,width=4cm]{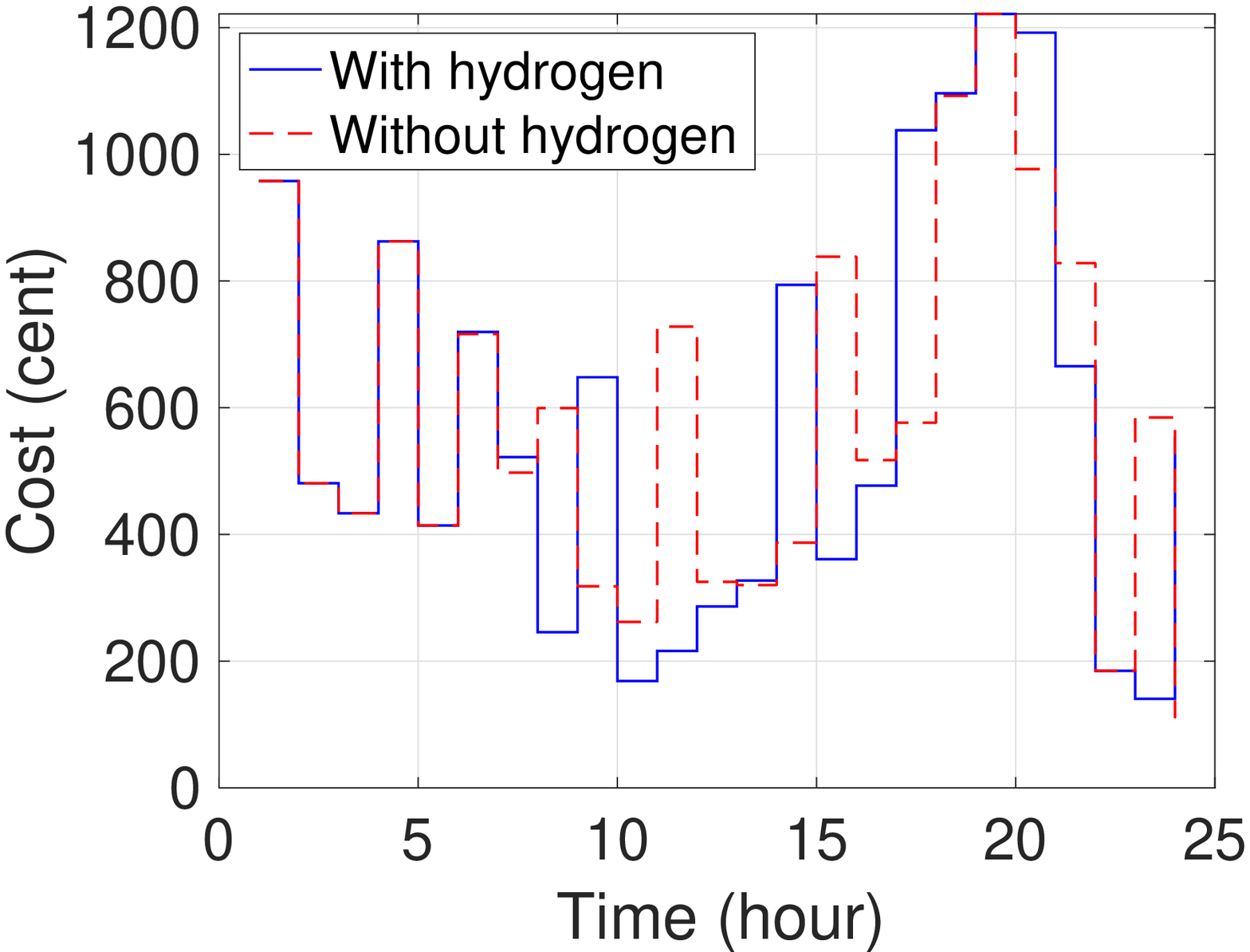}}
  \centerline{\scriptsize{(c) MG 3}}
\end{minipage}
\caption{Costs of each MG with and without hydrogen storage.}
\label{fig4}
\end{figure*}

\begin{figure*}
\centering
\begin{minipage}{0.33\linewidth}
  \centerline{\includegraphics[height=3cm,width=4cm]{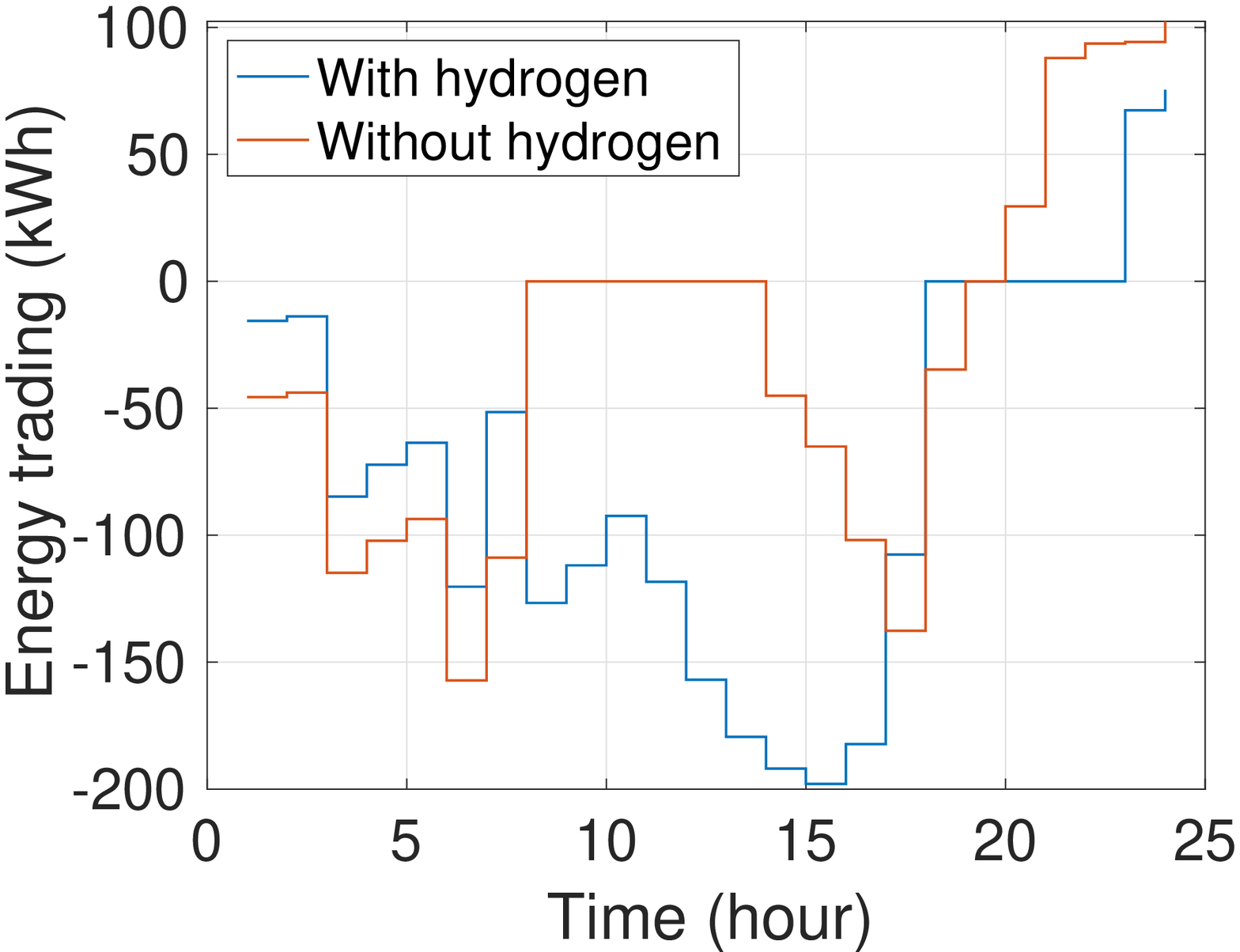}}
  \centerline{\scriptsize{(a) MG 1}}
\end{minipage}
\hspace{-5pt}
\begin{minipage}{0.33\linewidth}
  \centerline{\includegraphics[height=3cm,width=4cm]{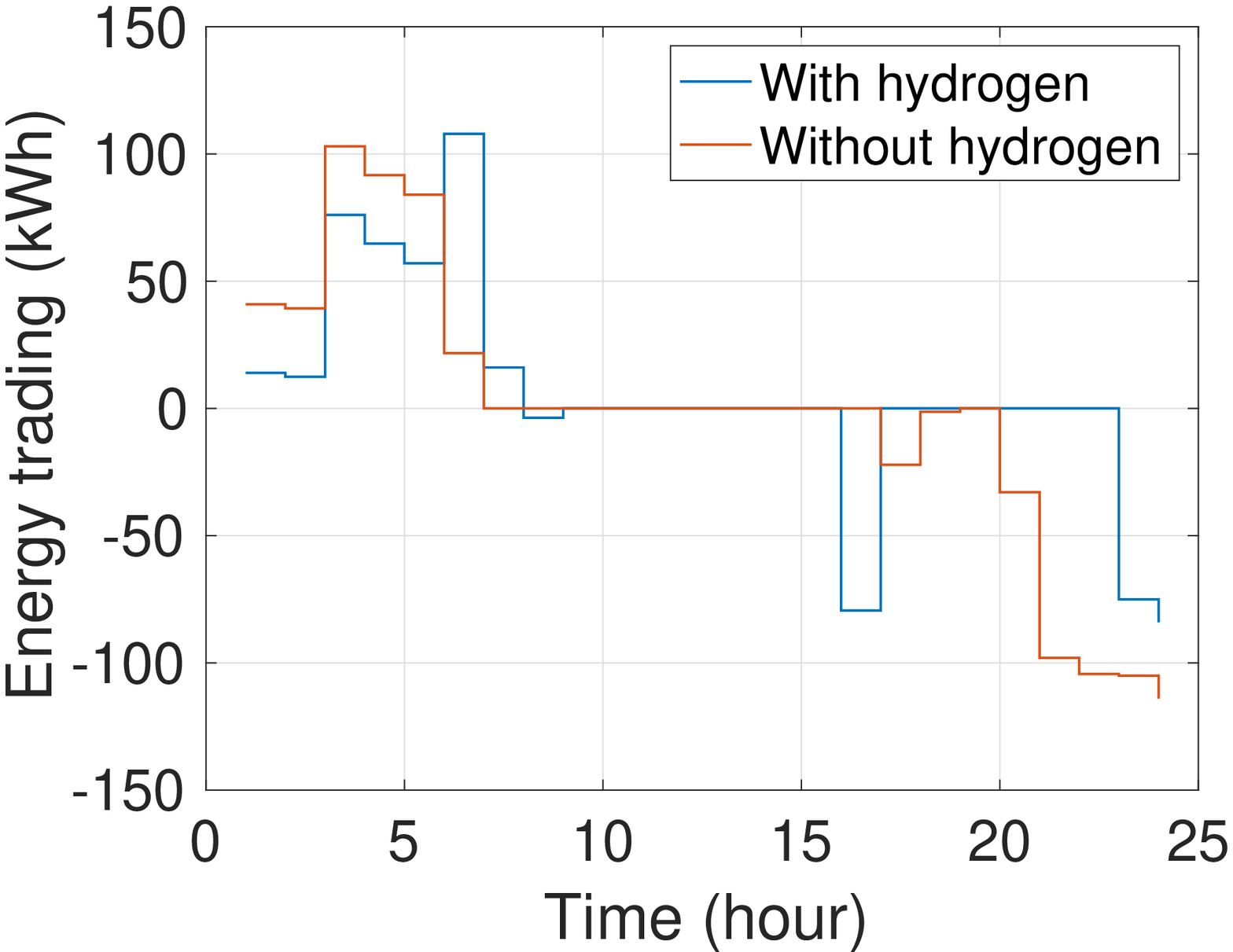}}
  \centerline{\scriptsize{(b) MG 2}}
\end{minipage}
\hspace{-5pt}
\begin{minipage}{0.33\linewidth}
  \centerline{\includegraphics[height=3cm,width=4cm]{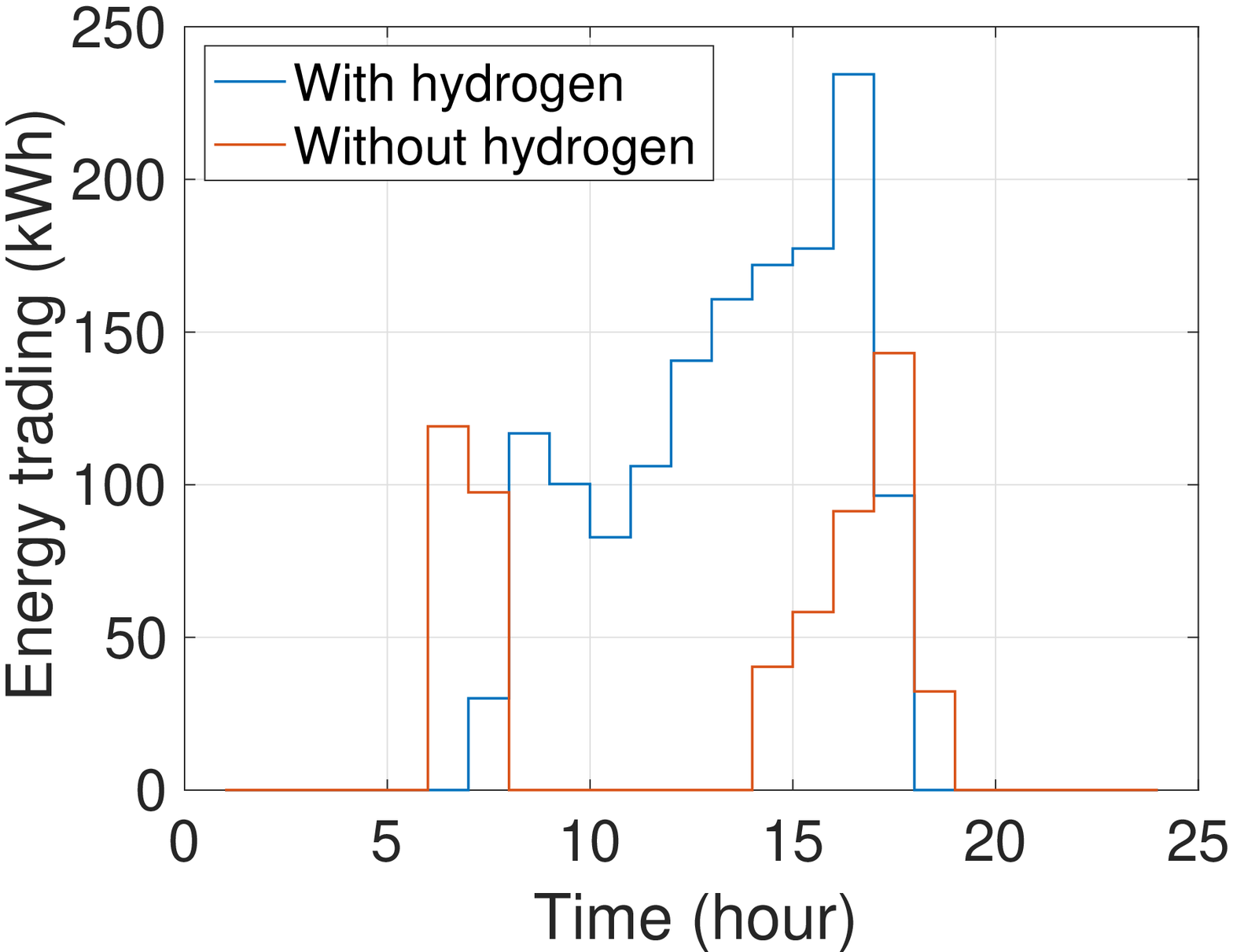}}
  \centerline{\scriptsize{(c) MG 3}}
\end{minipage}
\caption{Energy trading of each MG with and without hydrogen storage.}
\label{fig5}
\end{figure*}

\begin{figure*}
\centering
\begin{minipage}{0.33\linewidth}
  \centerline{\includegraphics[height=30mm,width=40mm]{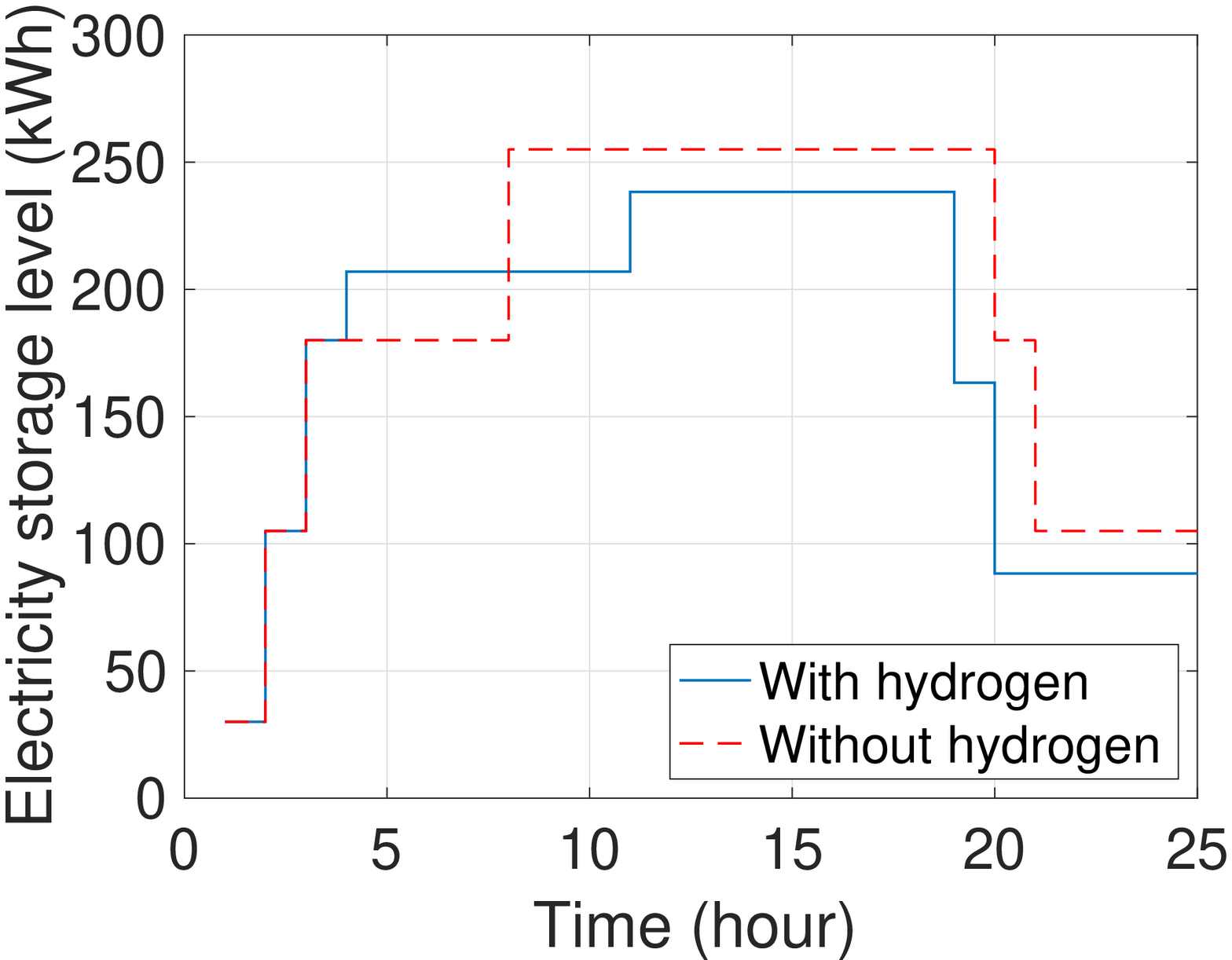}}
  \centerline{\scriptsize{(a) MG 1}}
\end{minipage}
\hspace{-5pt}
\begin{minipage}{0.33\linewidth}
  \centerline{\includegraphics[height=30mm,width=40mm]{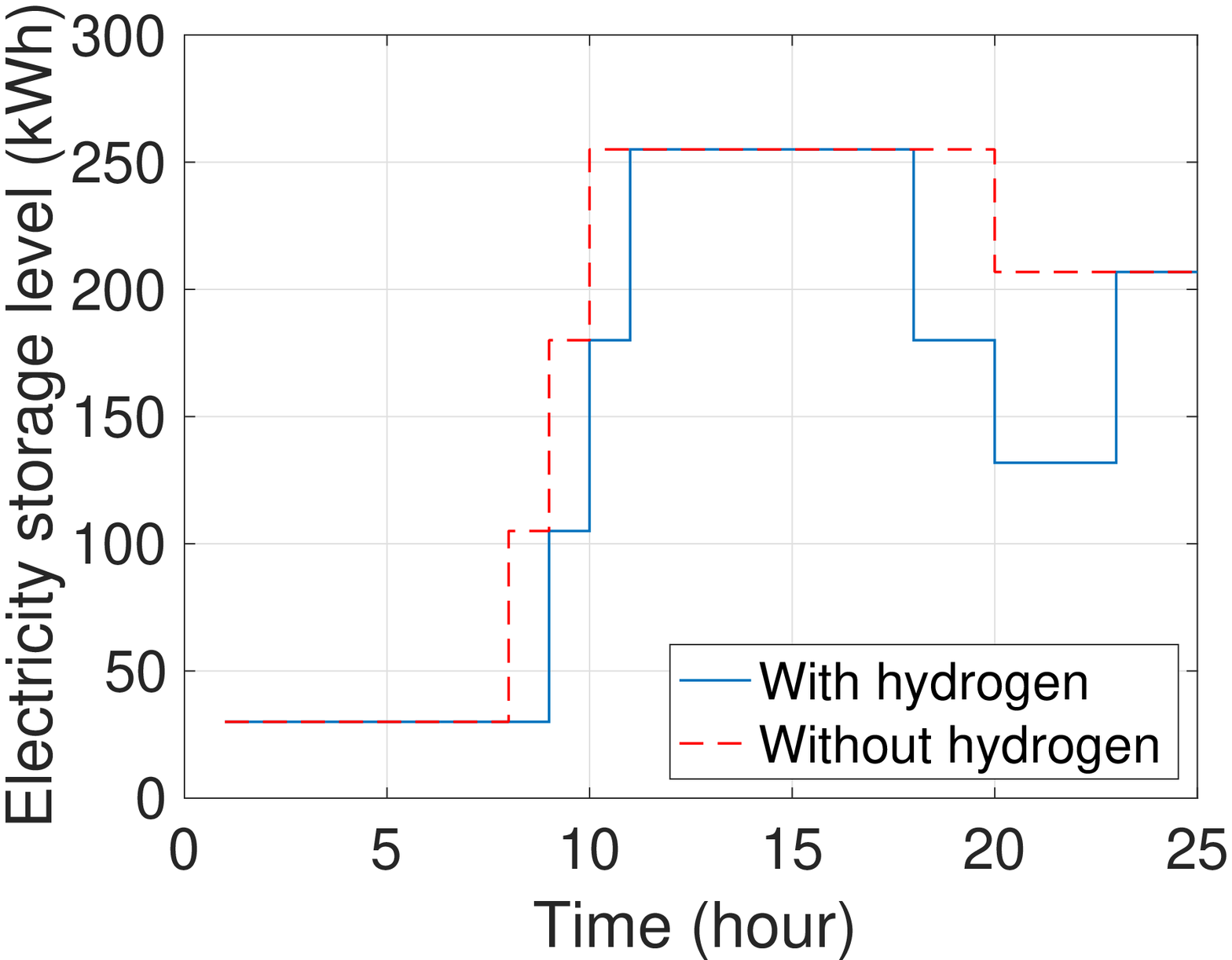}}
  \centerline{\scriptsize{(b) MG 2}}
\end{minipage}
\hspace{-5pt}
\begin{minipage}{0.33\linewidth}
  \centerline{\includegraphics[height=30mm,width=40mm]{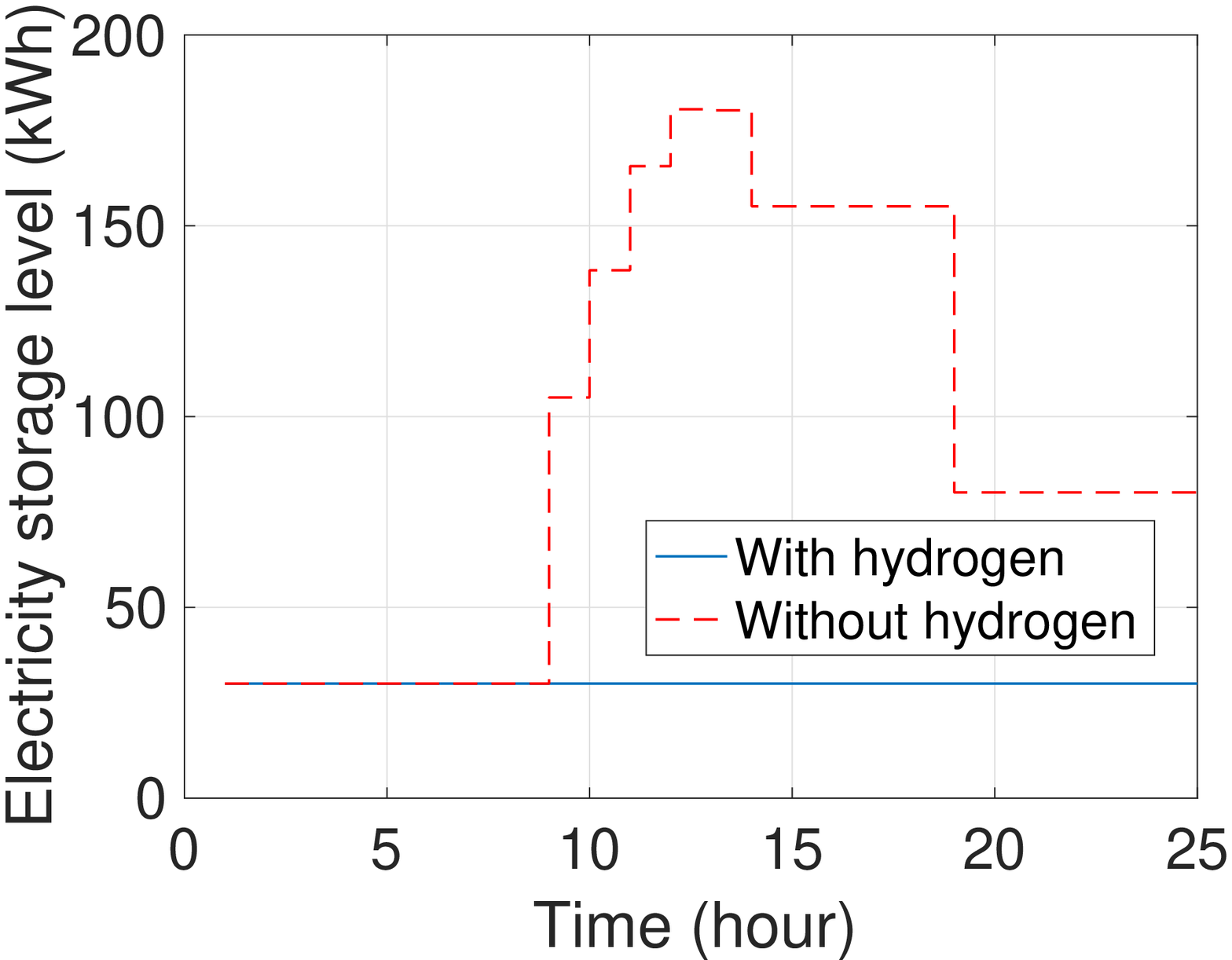}}
  \centerline{\scriptsize{(c) MG 3}}
\end{minipage}
\caption{Battery dynamics of each MG with and without hydrogen storage.}
\label{fig6}
\end{figure*}

Then, the comparisons of the costs, energy trading, and battery dynamics with and without hydrogen storage for all three MGs across 24 time slots are presented in Figs. \ref{fig4}-\ref{fig6}. During energy trading, positive values denote purchasing energy, and negative values denote selling energy. 
Fig. \ref{fig4} denotes that MGs achieve lower costs with hydrogen storage in most cases, where MGs electrolyze water to supply hydrogen for fuel cell vehicles or store hydrogen for the demand in the future instead of selling electricity to the electricity utility company at a low price. Fig. \ref{fig5} denotes the comparison of energy trading dynamics with and without hydrogen storage. MG $1$ sells more electricity to other MGs with hydrogen storage. This is because MG $3$ needs more electricity to electrolyze water to generate hydrogen for fuel cell vehicles, and MG $3$ purchases more electricity from MG $1$. Fig. \ref{fig6} denotes the comparison of battery dynamics with and without hydrogen storage. All MGs charge less electricity into the battery with hydrogen storage. This is because MGs with hydrogen storage use some electricity to electrolyze water to generate hydrogen.






\begin{figure*}
\centering
  \centerline{\includegraphics[height=42mm,width=56mm]{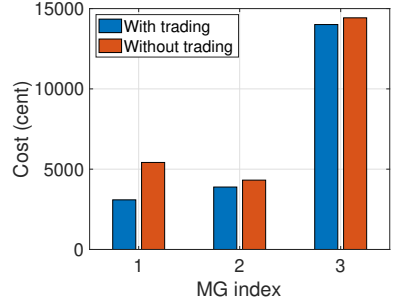}}
  \caption{Comparisons of all MGs' total costs with and without energy trading.}
  \label{fig7}
\end{figure*}

Energy trading plays an important role in releasing the imbalance of supply and demand for a single MG. Fig. \ref{fig7} shows that the costs of MGs are lower than those without trading. 
{
{Energy trading obviously reduces the cost of MG 1. The costs of MGs 2 and 3, however, are slightly reduced. The reason is that the renewable energy of MG 1 is more than the demand in most cases. MG 1 sells electricity to MGs 2 and 3 instead of selling electricity to the electricity company at a low price in most cases. Therefore, the cost of MG 1 is obviously reduced. MG 3 without energy trading cannot purchase electricity from MG 1 at a low price, but it can generate electricity by CHP system at a cost that is lower than the cost of purchasing electricity from the electricity company, and MG 2 trades little energy with other MGs. Therefore, the costs of MGs 2 and 3 with trading are slightly reduced in Fig. \ref{fig7}. }}

\begin{figure*}
\centering
\begin{minipage}{0.33\linewidth}
  \centerline{\includegraphics[height=30mm,width=40mm]{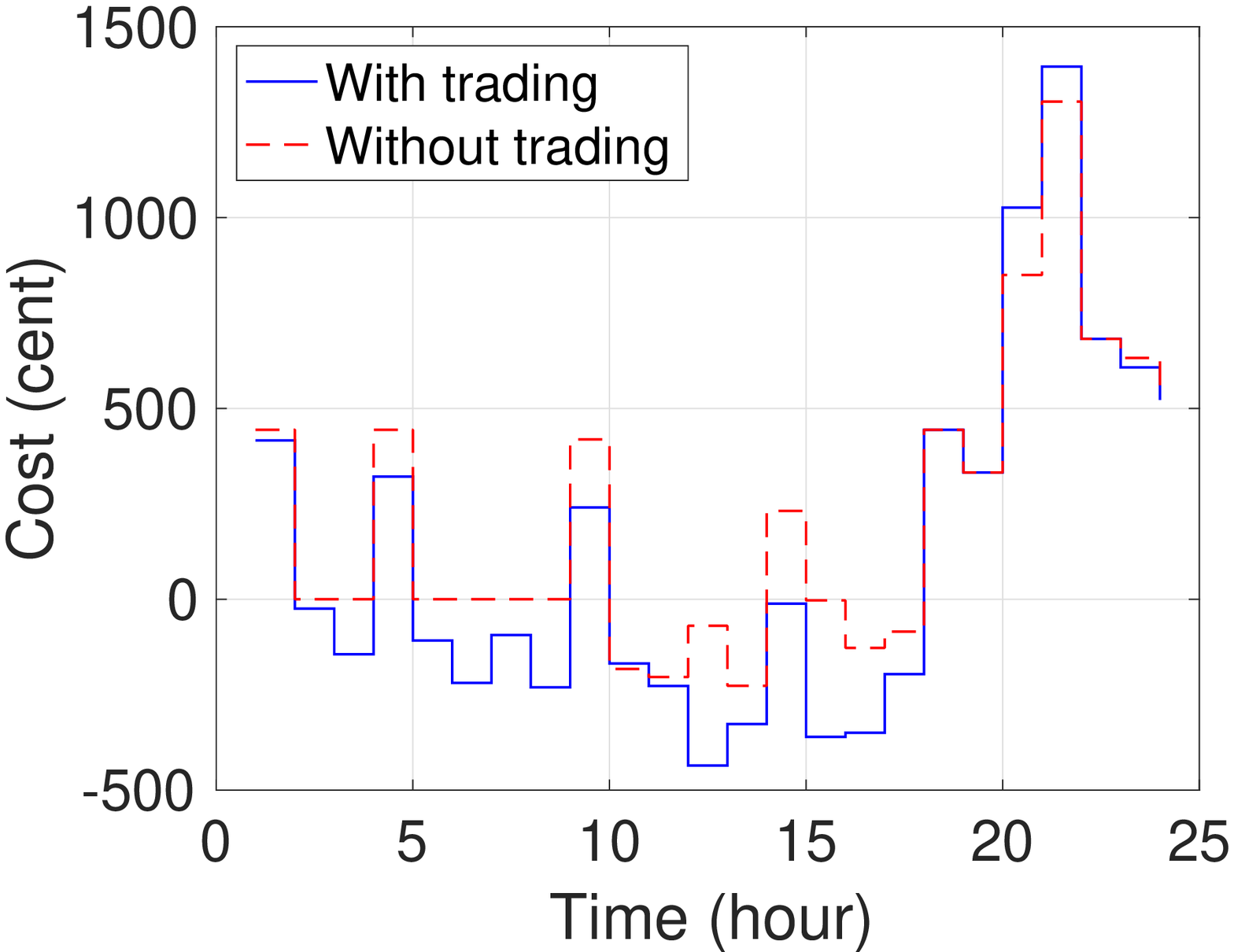}}
  \centerline{\scriptsize{(a) MG 1}}
\end{minipage}
\hspace{-5pt}
\begin{minipage}{0.33\linewidth}
  \centerline{\includegraphics[height=30mm,width=40mm]{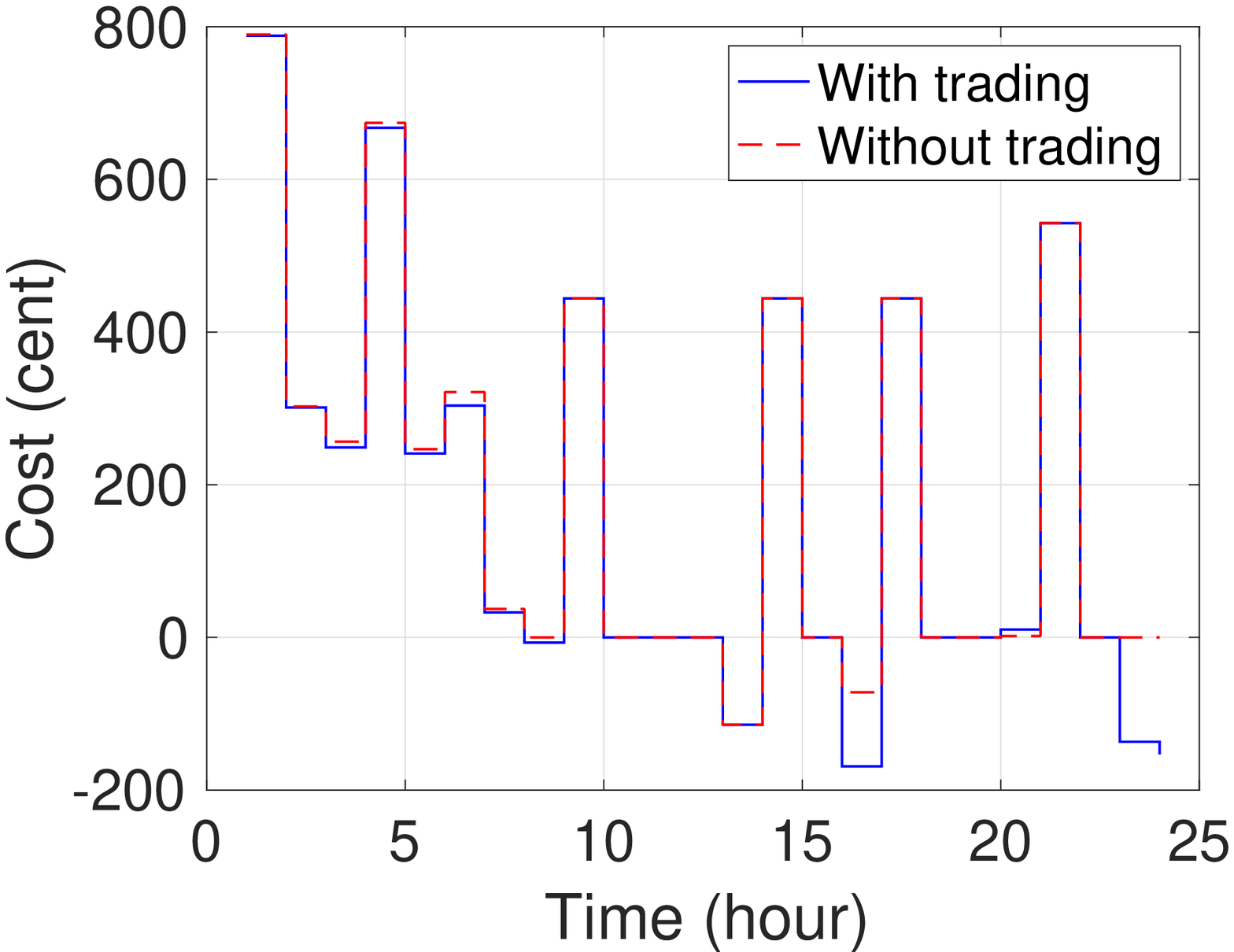}}
  \centerline{\scriptsize{(b) MG 2}}
\end{minipage}
\hspace{-5pt}
\begin{minipage}{0.33\linewidth}
  \centerline{\includegraphics[height=30mm,width=40mm]{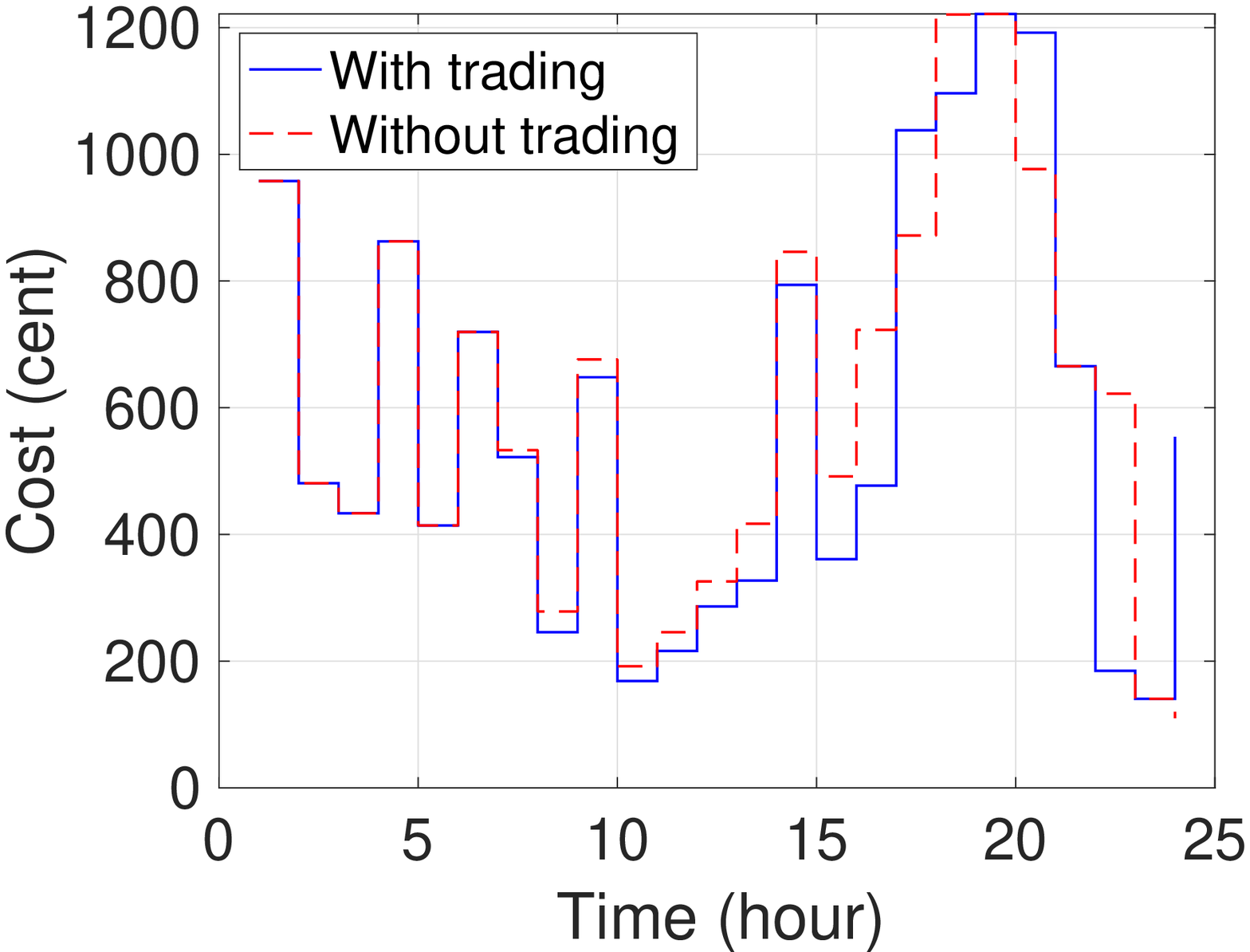}}
  \centerline{\scriptsize{(c) MG 3}}
\end{minipage}
\caption{Costs of each MG with and without energy trading.}
\label{fig8}
\end{figure*}

\begin{figure*}
\centering
\begin{minipage}{0.33\linewidth}
  \centerline{\includegraphics[height=30mm,width=40mm]{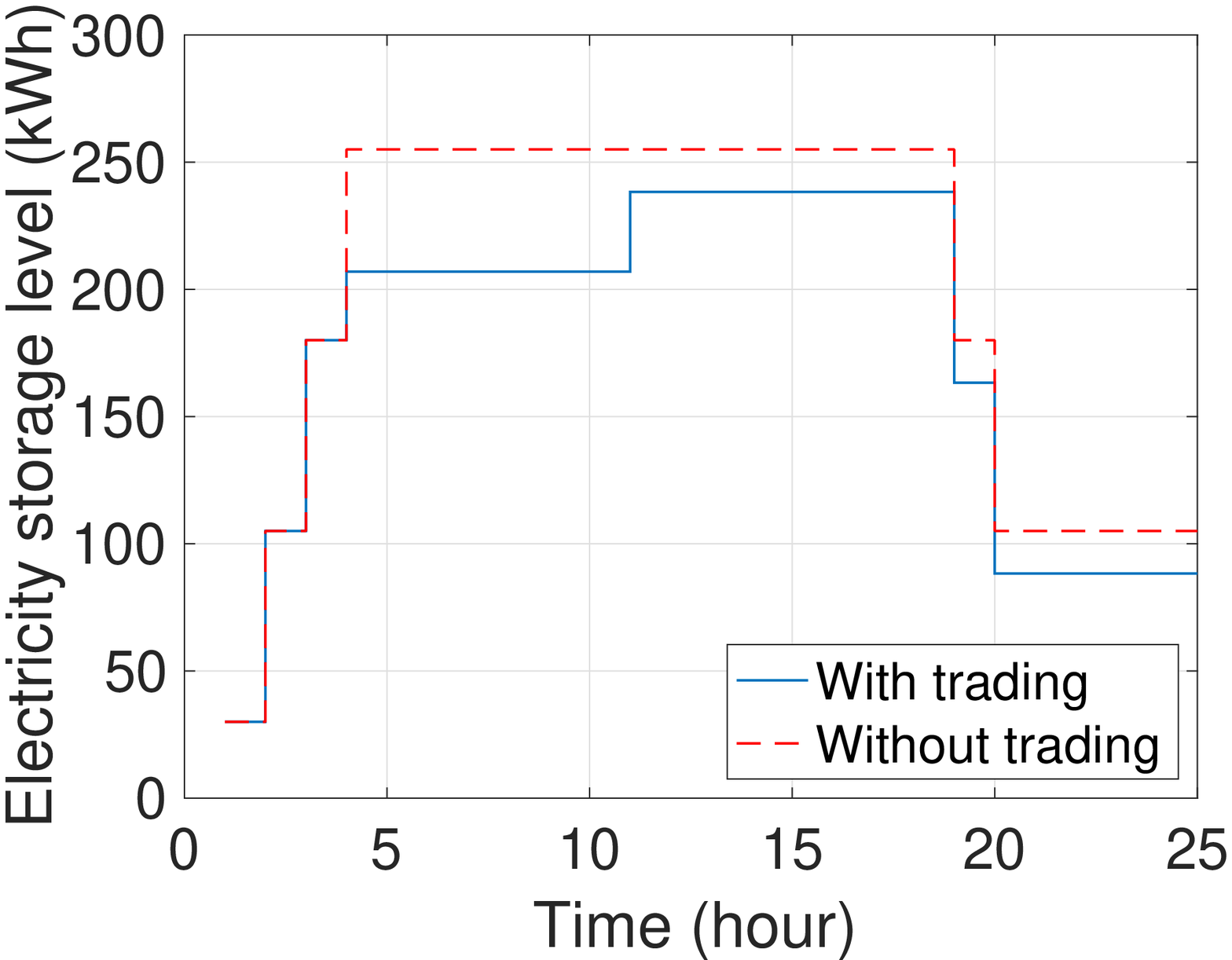}}
  \centerline{\scriptsize{(a) MG 1}}
\end{minipage}
\hspace{-5pt}
\begin{minipage}{0.33\linewidth}
  \centerline{\includegraphics[height=30mm,width=40mm]{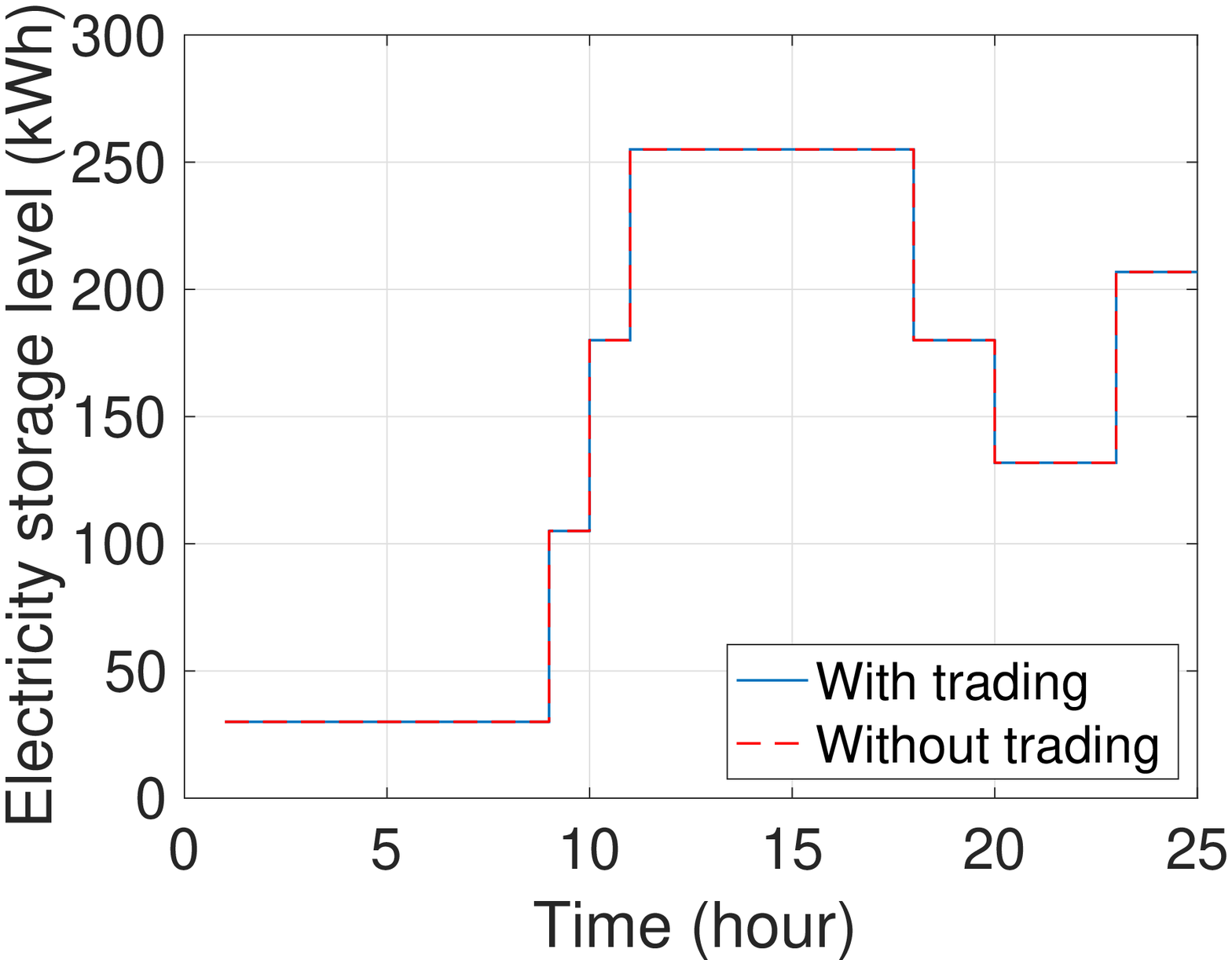}}
  \centerline{\scriptsize{(b) MG 2}}
\end{minipage}
\hspace{-5pt}
\begin{minipage}{0.33\linewidth}
  \centerline{\includegraphics[height=30mm,width=40mm]{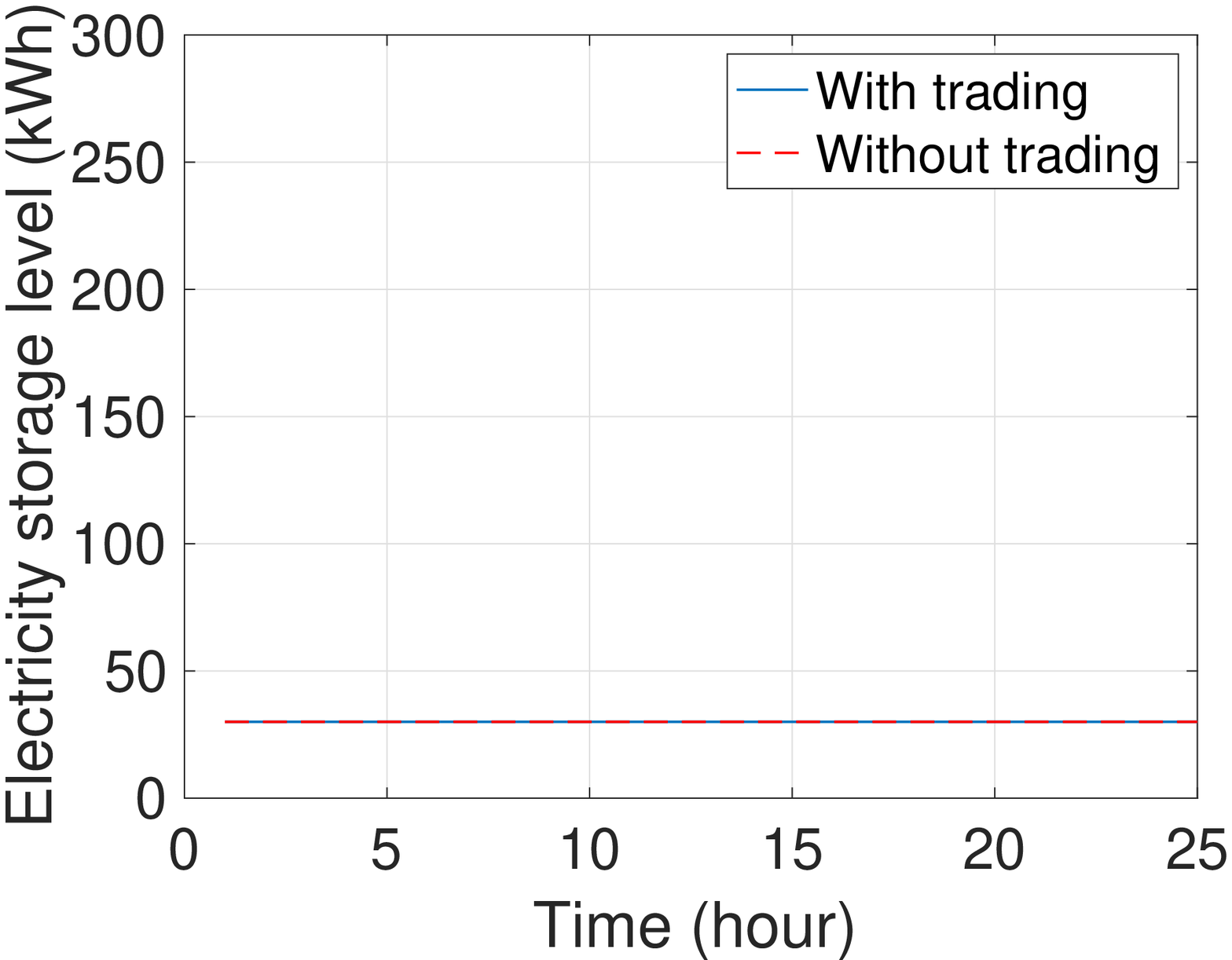}}
  \centerline{\scriptsize{(c) MG 3}}
\end{minipage}
\caption{Battery dynamics of each MG with and without energy trading.}
\label{fig9}
\end{figure*}

\begin{figure*}
\centering
\begin{minipage}{0.33\linewidth}
  \centerline{\includegraphics[height=30mm,width=40mm]{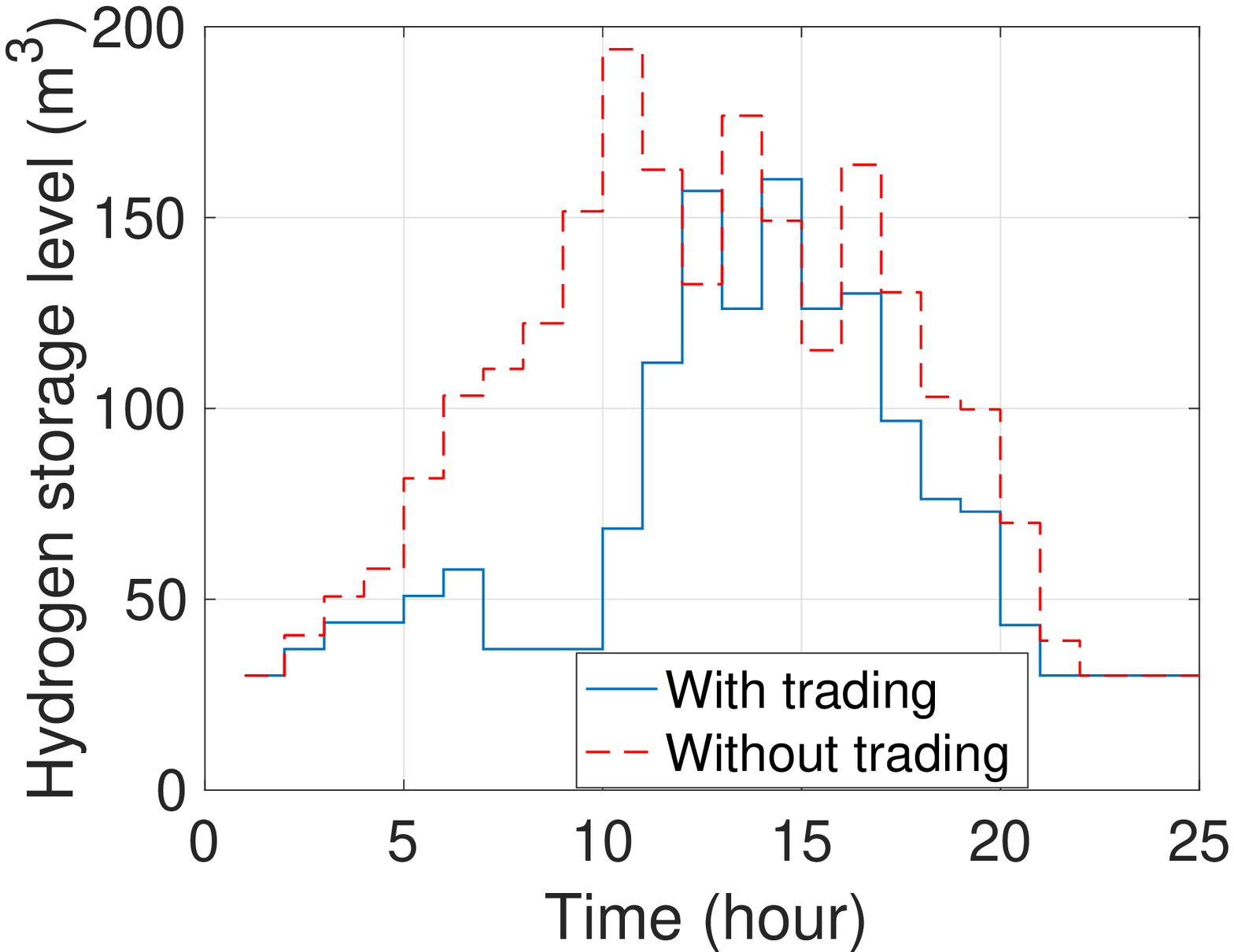}}
  \centerline{\scriptsize{(a) MG 1}}
\end{minipage}
\hspace{-5pt}
\begin{minipage}{0.33\linewidth}
  \centerline{\includegraphics[height=30mm,width=40mm]{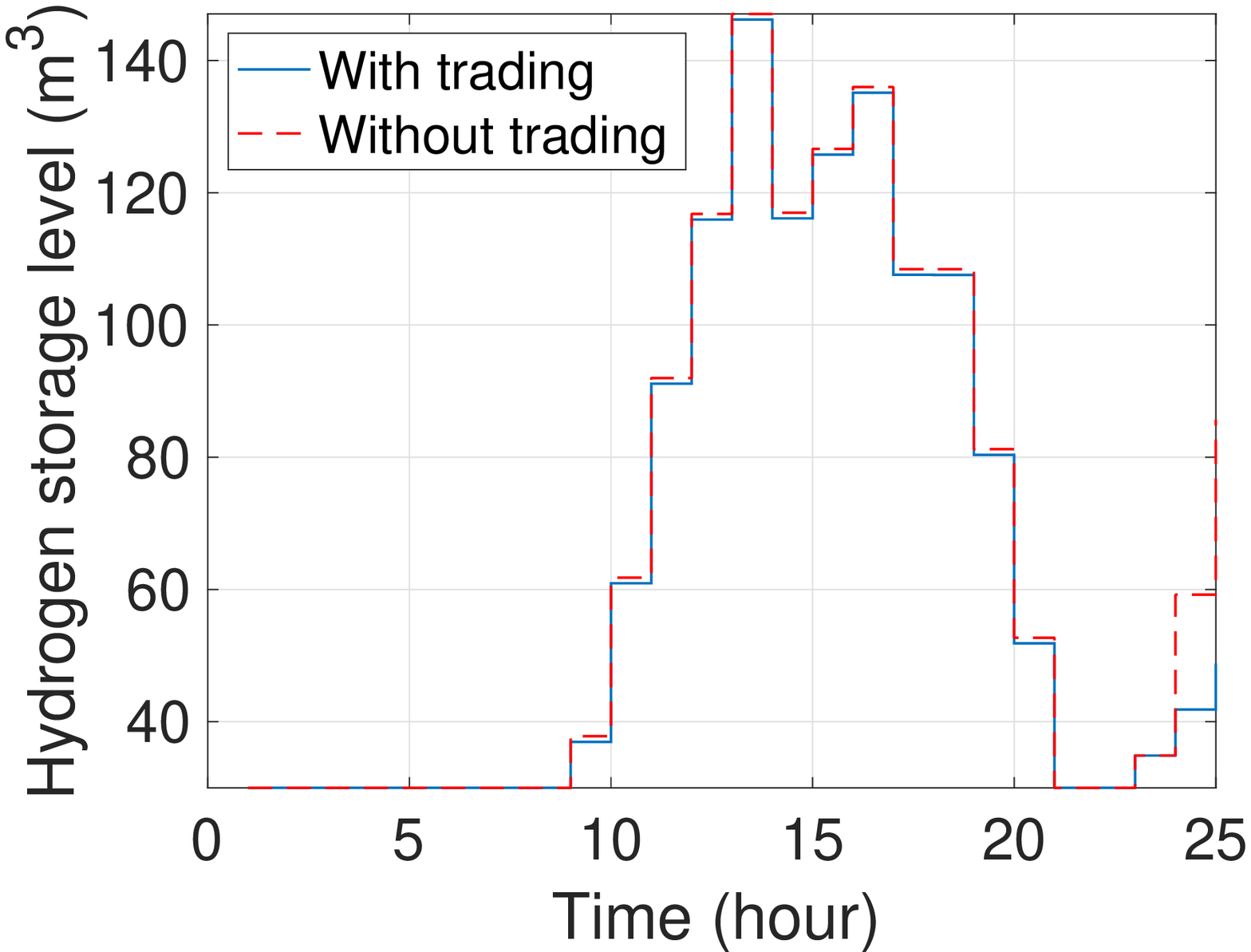}}
  \centerline{\scriptsize{(b) MG 2}}
\end{minipage}
\hspace{-5pt}
\begin{minipage}{0.33\linewidth}
  \centerline{\includegraphics[height=30mm,width=40mm]{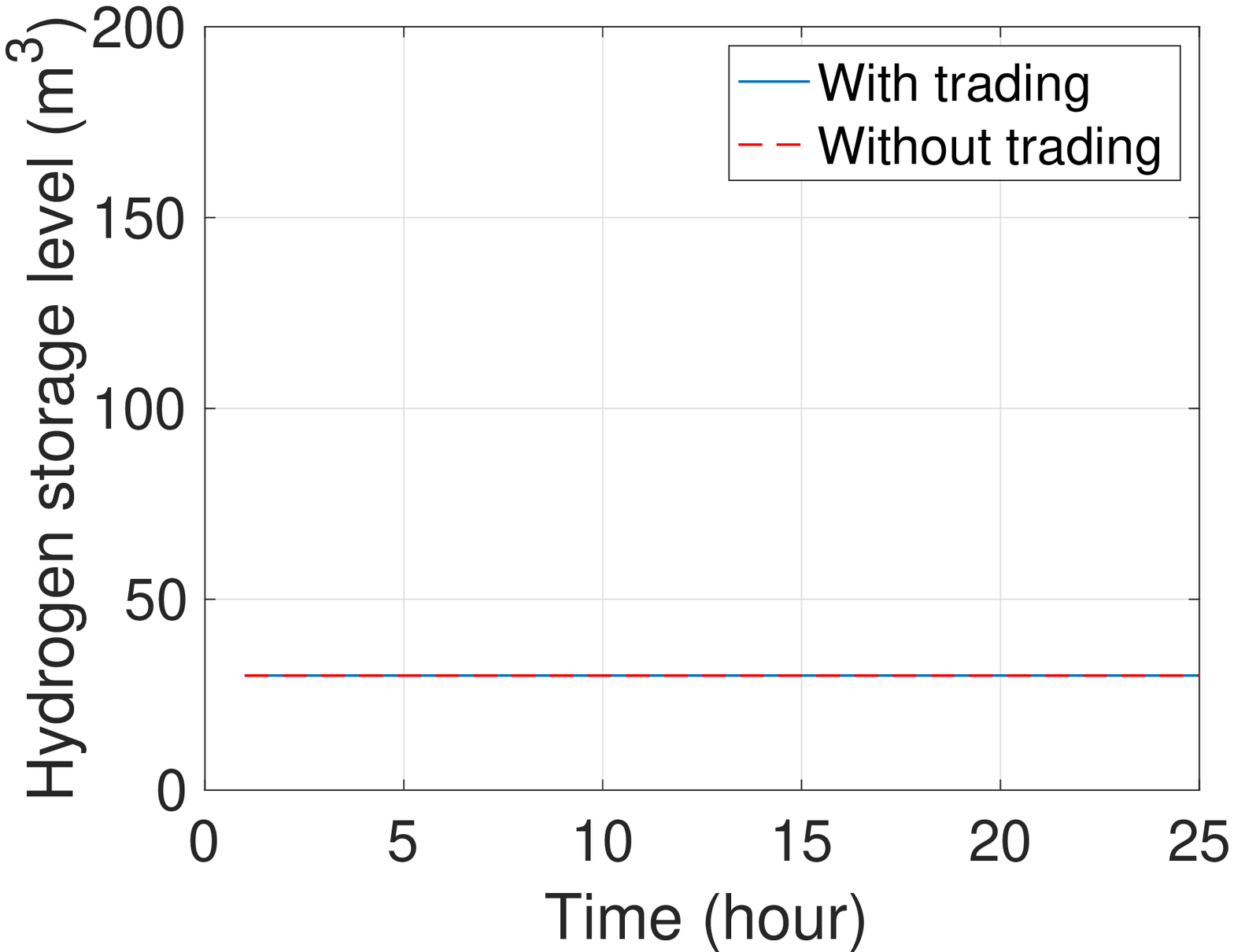}}
  \centerline{\scriptsize{(c) MG 3}}
\end{minipage}
\caption{Hydrogen storage dynamics of each MG with and without energy trading.}
\label{fig10}
\end{figure*}

Then, the comparisons of the costs, battery, and hydrogen storage dynamics with and without energy trading for all three MGs across 24 time slots are presented in Figs. \ref{fig8}-\ref{fig10}. 
Fig. \ref{fig8} shows that MGs achieve lower costs with energy trading in most cases, where MGs acquire electricity from other MGs in energy trading instead of the electricity utility company. Fig. \ref{fig2}(b) tells that MG $1$ has higher renewable energy output than other MGs, and hence MG $1$ sells excessive energy to other MGs in most cases except the last four hours. The reason is because MG $1$ has a drop in renewable energy output during the last four hours, while MG $2$ has adequate renewable energy output during the last four hours. Fig. \ref{fig9} denotes the comparison of battery dynamics with and without energy trading. MG $1$ charges less electricity into the battery with energy trading. This is because MG $1$ sells electricity to other MGs instead of storing electricity in the battery. This is the same as the hydrogen storage in Fig. \ref{fig10}. 
{
{Whether it involves trading or not, MG 3 without abundant renewable energy to electrolyze water, has to purchase hydrogen from a hydrogen-producing company to supply vehicles. Therefore, MG 3 has no energy to charge, and the dynamics of the storage level of MG 3 are the same as in Fig. \ref{fig9} and Fig. \ref{fig10}.}}

Table 1 shows the costs of three MGs under the proposed method, and the methods without hydrogen storage and without energy trading. {
{Three cases of different initial energy storages are studied as follows. 1) The initial energy of storage is 10\% of its capacity (Figs. \ref{fig3}-\ref{fig10} are generated in this case). The total cost of three MGs is reduced by up to 26.53\% from 28563 cents without hydrogen storage to 20984 cents, and 13.16\% from 24163 cents without energy trading to 20984 cents. 
2) The initial energy of storage is 50\% of its capacity. The total cost of three MGs is reduced by up to 29.68\% from 26121 cents without hydrogen storage to 18367 cents, and 15.92\% from 21844 cents without energy trading to 18367 cents. 
3) The initial energy of storage is its capacity. The total cost of three MGs is reduced by up to 35.50\% from 22888 cents without hydrogen storage to 14763 cents, and 19.55\% from 18350 cents without energy trading to 14763 cents. 
According to the results, it is observed that the cost decreases with an increase in the initial energy of storage, and the extent of cost reduction increases with an increase in the initial energy of storage. This is because more initial energy of storage means less cost of purchasing energy and more energy used to trade or electrolyze water.}}
The introduction of hydrogen storage and energy trading reduces the costs of all MGs. 
Therefore, MGs benefit from energy trading, hydrogen storage, and fuel cell vehicles. This verifies the effectiveness of the proposed algorithm.

\begin{table}  
\small  
\caption{{
{Costs of three MGs under different methods.}}}  
\begin{center}  
\begin{tabular}{l|l|l|l|l|l}  
\hline  
Initial energy storage&Costs (cent)  & MG $1$ & MG $2$ & MG $3$ & System \\ \hline  
\multirow{3}*{10\% of capacity}&Cost & 3091 & 3887 & 14006 & 20984 \\   
\cline{2-6} 
{}&Cost(without trading) & 5419 & 4319 & 14425 & 24163 \\  
\cline{2-6} 
{}&Cost(without hydrogen) & 7327 & 7005 & 14231 & 28563 \\ \hline  
\multirow{3}*{50\% of capacity}&Cost & 2060 & 3377 & 12931 & 18367 \\ \cline{2-6}  
{}&Cost(without trading) & 4570 & 3738 & 13536 & 21844 \\ \cline{2-6}   
{}&Cost(without hydrogen) & 6392 & 6491 & 13239 & 26121 \\ \hline  
\multirow{3}*{100\% of capacity}&Cost & 807 & 2198 & 11758 & 14763 \\ \cline{2-6}   
{}&Cost(without trading) & 3419 & 2567 & 12364 & 18350 \\ \cline{2-6}   
{}&Cost(without hydrogen) & 5509 & 5317 & 12061 & 22888\\ \hline  
\end{tabular}  
\end{center}  
\end{table}

\section{Conclusion}
In this paper, the energy scheduling and energy trading problem for real-time pricing among multiple microgrids is studied, which is an urgent issue for today's the cyber-physical-energy systems. A multi-energy management framework including fuel cell vehicles, energy storage, {
{combined heat and power system}}, and renewable energy is presented, where fuel cell vehicles and energy storage further improve the absorption of the renewable energy. 
A joint algorithm based on Lyapunov optimization and a double-auction mechanism is designed to optimize the long-term energy cost of each microgrid. 
At last, the results based on real data show that microgrids' costs, under the management of the proposed algorithm, can be decreased. 
Comparative analysis of energy storage and energy trading demonstrate the necessity of including energy storage and trading.

In this paper, fuel cell vehicles refer to buses that have a specific route. Due to transportation concerns, fuel cell vehicles can be cars, buses, and so on. In this case, the trip characteristics of vehicles need to be considered. Investigating some control schemes, e.g., Ref. \cite{AlaviFuel}, to optimize dispatch of fuel cell vehicles is a significant research direction. Another direction is how to design the scheduling method, e.g., Ref. \cite{ZhouIndirect}, to further realize the multi-energy coupled peak load shifting in realistic scenarios, such as industrial parks.


\begin{appendix}
\section{Proof of (\ref{rightmin})  }
According to (\ref{A1}) - (\ref{A3}), (\ref{Yc}), and (\ref{vir}), the Lyapunov drift term $\Delta_{i}(t)$ is denoted by
\begin{equation}
\begin{aligned}
&\Delta_{i}(t)=\mathbb{E}\{Q_{i}(t+1)-Q_{i}(t)|B_{i}(t),Y_{i}(t),W_{i}(t),Y_{il}(t)\}
\\&=\frac{1}{2}\mathbb{E}\{2A_{i}(t)(C_{ie}(t)-D_{ie}(t))+2F_{i}(t)(C_{iy}(t)-D_{iy}(t))\\&+2Z_{i}(t)(C_{ih}(t)-D_{ih}(t))+\sum^{L_i}_{l=1}[2I_{il}(t)(D_{iyl}(t)+d_{il}(t)\\&-Y_{ifl}(t)-h_{il}(t))]+(C_{ie}(t)-D_{ie}(t))^2\\&+(C_{iy}(t)-D_{iy}(t))^2+(C_{ih}(t)-D_{ih}(t))^2\\&+\sum^{L_i}_{l=1}[(D_{iyl}(t)+d_{il}(t)-Y_{ifl}(t)-h_{il}(t))^2]\}
\\& \leq \frac{1}{2}\mathbb{E}\{2A_{i}(t)(C_{ie}(t)-D_{ie}(t))+2F_{i}(t)(C_{iy}(t)-D_{iy}(t))\\&+2Z_{i}(t)(C_{ih}(t)-D_{ih}(t))+\sum^{L_i}_{l=1}[2I_{il}(t)(D_{iyl}(t)+d_{il}(t)\\&-Y_{ifl}(t)-h_{il}(t))]+\max(C_{ie,max}^2,D_{ie,max}^2)\\&+\max(C_{iy,max}^2,D_{iy,max}^2)+\max(C_{ih,max}^2,D_{ih,max}^2)\\&+\sum^{L_i}_{l=1}[\max((D_{iyl,max}+d_{il,max})^2,(Y_{ifl,max}+h_{il,max})^2)]\}
\\&=\mathbb{E}\{A_{i}(t)(C_{ie}(t)-D_{ie}(t))+F_{i}(t)(C_{iy}(t)-D_{iy}(t))\\&+Z_{i}(t)(C_{ih}(t)-D_{ih}(t))+\sum^{L_i}_{l=1}[I_{il}(t)(D_{iyl}(t)+d_{il}(t)\\&-Y_{ifl}(t)-h_{il}(t))]\}+G_{i}
\end{aligned}
\end{equation}
where $G_{i}=\frac{1}{2}\{\max(C_{ie,max}^2,D_{ie,max}^2)+\max(C_{iy,max}^2,D_{iy,max}^2)+\max(C_{ih,max}^2,D_{ih,max}^2)+\sum^{L_i}_{l=1}[\max((D_{iyl,max}+d_{il,max})^2,(Y_{ifl,max}+h_{il,max})^2)]\}$

\section{Proof of Lemma 1 }
The following four cases are considered to determine the price of energy trading:
\begin{enumerate}
\item Case 1: $A_{i}(t)\geq0$. In this case, MG $i$ has too much energy in its battery. According to (\ref{lg}),
$C_{ie}(t)-D_{ie}(t)=E_{i}(t)+N_{i}(t)+X_{i}(t)-S_{i}(t)
+\eta_fhY_{if}(t)-\frac{hC_{iy}(t)}{\eta_e}-c_1C_{iy}(t)+\eta_{pg}P_{i}^{CHP}(t)-E_{io}(t)-L_{ie}(t)$. According to (\ref{P2}), i.e.
\begin{equation}
\begin{aligned}
&\min_{\boldsymbol{M}_{i}(t)} A_{i}(t)(C_{ie}(t)-D_{ie}(t))+F_{i}(t)(C_{iy}(t)-D_{iy}(t))\\&+Z_{i}(t)(C_{ih}(t)-D_{ih}(t))+\sum^{L_i}_{l=1}[I_{il}(t)(D_{iyl}(t)+d_{il}(t)\\&-Y_{ifl}(t)-h_{il}(t))]
+V_{i}(
p_{y}(t)d_{i}(t)
+E_{i}(t)p_{e}(t)\\&+ (P_{i}^{CHP}(t)+H_{i}^{CHP}(t)+H_{i}^b(t))p_{g}(t)+\beta_i(t) X_{i}(t)\\&-\alpha_i(t) S_{i}(t)-E_{io}(t)p_{eo}(t))\\
=&\min_{\boldsymbol{M}_{i}(t)} -(A_{i}(t)+V_{i}\alpha_i(t))S_{i}(t)+(A_{i}(t)+V_{i}\beta_i(t))X_{i}(t)\\&+A_{i}(t)(E_{i}(t)+N_{i}(t)+\eta_fhY_{if}(t)-\frac{hC_{iy}(t)}{\eta_e}\\&-c_1C_{iy}(t)+\eta_{pg}P_{i}^{CHP}(t)-E_{io}(t)-L_{ie}(t))\\&+F_{i}(t)(C_{iy}(t)-D_{iy}(t))+Z_{i}(t)(C_{ih}(t)-D_{ih}(t))\\&+\sum^{L_i}_{l=1}[I_{il}(t)(D_{iyl}(t)+d_{il}(t)-Y_{ifl}(t)-h_{il}(t))]
\\&+V_{i}(p_{y}(t)d_{i}(t)+E_{i}(t)p_{e}(t)+ (P_{i}^{CHP}(t)+H_{i}^{CHP}(t)\\&+H_{i}^b(t))p_{g}(t)-E_{io}(t)p_{eo}(t))
\end{aligned}
\label{cde}
\end{equation}

and $-\alpha_i(t)V_{i}-A_{i}(t)<0$, MG $i$ tends to increase $S_i(t)$, and $C_{ie}(t)=0$, $D_{ie}(t)=D_{ie,max}$.

\item Case 2: $A_{i}(t)<0$. In this case, six situations are considered.
\begin{itemize}
\item If $0<\alpha_i(t)<\frac{-A_{i}(t)}{V_{i}}$, then$-\alpha_i(t)V_{i}-A_{i}(t)>0$. Therefore, MG $i$ tends to decrease $S_i(t)$ and increase $C_{ie}(t)$.
\item If $\alpha_i(t)>\frac{-A_{i}(t)}{V_{i}}$, then $-A_{i}(t)-\alpha_i(t)V_{i}<0$. Therefore, MG $i$ tends to increase $S_i(t)$ and decrease $C_{ie}(t)$. 
\item If $\alpha_i(t)=\frac{-A_{i}(t)}{V_{i}}$, then $-A_{i}(t)-\alpha_i(t)V_{i}=0$. This is the same for MG $i$ to increase $S_i(t)$ or increase $C_{ie}(t)$. 

\item If $0<\beta_i(t)<\frac{-A_{i}(t)}{V_{i}}$, then $\beta_i(t)V_{i}+A_{i}(t)<0$. Therefore, MG $i$ tends to increase $X_i(t)$ and decrease $D_{ie}(t)$.
\item If $\beta_i(t)>\frac{-A_{i}(t)}{V_{i}}$, then $A_{i}(t)+\beta_i(t)V_{i}>0$. Therefore, MG $i$ tends to decrease $X_i(t)$ and increase $D_{ie}(t)$.
\item If $\beta_i(t)=\frac{-A_{i}(t)}{V_{i}}$, then $-A_{i}(t)=\beta_i(t)V_{i}$. It is same for MG $i$ to increase $X_i(t)$ or increase $D_{ie}(t)$. 
\end{itemize}
 Case 3: $F_{i}(t)\geq0$. In this case, MG $i$ has too much hydrogen in its hydrogen storage. According to (\ref{lg}), $(\frac{h}{\eta_e}+c_1)C_{iy}(t)=E_{i}(t)+N_{i}(t)+X_{i}(t)-S_{i}(t)-C_{ie}(t)+D_{ie}(t)+\eta_fhY_{if}(t)+\eta_{pg}P_{i}^{CHP}(t)-E_{io}(t)-L_{ie}(t)$. According to (\ref{P2}), i.e.
\begin{equation}
\begin{aligned}
&\min_{\boldsymbol{M}_{i}(t)} A_{i}(t)(C_{ie}(t)-D_{ie}(t))+F_{i}(t)(C_{iy}(t)-D_{iy}(t))\\&+Z_{i}(t)(C_{ih}(t)-D_{ih}(t))+\sum^{L_i}_{l=1}[I_{il}(t)(D_{iyl}(t)+d_{il}(t)\\&-Y_{ifl}(t)-h_{il}(t))]+V_{i}(p_{y}(t)d_{i}(t)
+E_{i}(t)p_{e}(t)\\&+ (P_{i}^{CHP}(t)+H_{i}^{CHP}(t)+H_{i}^b(t))p_{g}(t)+\beta_i(t) X_{i}(t)\\&-\alpha_i(t) S_{i}(t)-E_{io}(t)p_{eo}(t))\\
=&\min_{\boldsymbol{M}_{i}(t)} A_{i}(t)(C_{ie}(t)-D_{ie}(t))-(\frac{F_{i}(t)}{\frac{h}{\eta_e}+c_1}+V_i\alpha_i(t))S_{i}(t)\\&+(\frac{F_{i}(t)}{\frac{h}{\eta_e}+c_1}+V_i\beta_i(t))X_{i}(t)+\frac{F_{i}(t)}{\frac{h}{\eta_e}+c_1}(E_{i}(t)+N_{i}(t)\\&-C_{ie}(t)+D_{ie}(t)+\eta_fhY_{if}(t)+\eta_{pg}P_{i}^{CHP}(t)-E_{io}(t)\\&-L_{ie}(t))-F_{i}(t)D_{iy}(t)+Z_{i}(t)(C_{ih}(t)-D_{ih}(t))\\&+\sum^{L_i}_{l=1}[I_{il}(t)(D_{iyl}(t)+d_{il}(t)-Y_{ifl}(t)-h_{il}(t))]\\&+V_{i}(p_{y}(t)d_{i}(t)
+E_{i}(t)p_{e}(t)-E_{io}(t)p_{eo}(t)\\&+ (P_{i}^{CHP}(t)+H_{i}^{CHP}(t)+H_{i}^b(t))p_{g}(t))
\end{aligned}
\label{cdh}
\end{equation} and $-\alpha_i(t)V_{i}-\frac{F_{i}(t)}{\frac{h}{\eta_e}+c_1}<0$, MG $i$ tends to increase $S_i(t)$, and $C_{iy}(t)=0$, $D_{iy}(t)=D_{iy,max}$.

\item Case 4: $F_{i}(t)<0$. In this case, three situations are considered.
\begin{itemize}
\item If $0<\alpha_i(t)<\frac{-F_{i}(t)}{(\frac{h}{\eta_{e}}+c_1)V_{i}}$, then $-\alpha_i(t)V_{i}-\frac{F_{i}(t)}{\frac{h}{\eta_e}+c_1}>0$. Therefore, MG $i$ tends to decrease $S_i(t)$ and increase $C_{iy}(t)$. 
\item If $\alpha_i(t)>\frac{-F_{i}(t)}{(\frac{h}{\eta_{e}}+c_1)V_{i}}$, then $-\frac{F_{i}(t)}{\frac{h}{\eta_e}+c_1}-\alpha_i(t)V_{i}<0$. Therefore, MG $i$ tends to increase $S_i(t)$ and decrease $C_{iy}(t)$. 
\item If $\alpha_i(t)=\frac{-F_{i}(t)}{(\frac{h}{\eta_{e}}+c_1)V_{i}}$, then $-\frac{F_{i}(t)}{\frac{h}{\eta_e}+c_1}-\alpha_i(t)V_{i}=0$. It is same for MG $i$ to increase $S_i(t)$ or increase $C_{iy}(t)$. 
\end{itemize}
\end{enumerate}

\section
{Proof of Lemma 2  }
All MGs are assumed to be rational. They will choose a strategy that can minimize their costs. The purchase price and selling price submitted by MG $i$ are $\beta_{i}(t)$ and $\alpha_{i}(t)$, and the purchase price and selling price determined by the double-auction mechanism are $\hat{\beta}(t)$ and $\hat{\alpha}(t)$ in the actual energy trading. Then, the benefit of MG $i$ is analyzed when it cheats. 
\begin{enumerate}
\item Case 1: $\alpha_{i}(t) > \hat{\alpha}(t)$. In this case, MG $i$ is not allowed to sell energy in the double-auction mechanism. 

\begin{itemize}
\item 
If MG $i$ increases $\alpha_{i}(t)$, the situation does not change. \item 
If MG $i$ reduces $\alpha_{i}(t)$ and $\alpha_{i}(t) > \hat{\alpha}(t)$, the situation does not change.
\item If MG $i$ reduces $\alpha_{i}(t)$ and $ \alpha_{i}(t) \leq \hat{\alpha}(t)$, the MG will be forced to sell energy at a price lower than expected, and its benefit will decrease owing to cheating. 
\end{itemize}

\item Case 2: $\alpha_{i}(t) \leq \hat{\alpha}(t)$. In this case, MG $i$ sells energy in the double-auction mechanism. 
\begin{itemize}
\item
If MG $i$ reduces $\alpha_{i}(t)$, the situation does not change. 
\item
If MG $i$ increases $\alpha_{i}(t)$ and $ \alpha_{i}(t) \leq \hat{\alpha}(t)$, the situation does not change. 
\item If MG $i$ increases $\alpha_{i}(t)$ and $\alpha_{i}(t) > \hat{\alpha}(t)$, the MG is not allowed to sell energy in the double-auction mechanism. However, energy is excessive, MG $i$ may sell excessive energy to the electricity utility company at a lower price, and its benefit will decrease owing to cheating. 
\end{itemize}

\end{enumerate}

This is similar to analyze $\beta_{i}(t)$. Therefore, the double-auction mechanism can prevent MGs from cheating.

\section{Proof of Lemma 3  }
The induction is used to prove the bound of $B_{i}(t)$, $Y_{i}(t)$, $W_{i}(t)$, and $Y_{il}(t)$. First, the conditions hold at time slot 1 and still hold at time slot $t$. Then, the following four cases are considered.
\begin{enumerate}
\item Case 1: $B_{i}(t)<\theta_{i}$. In this case, $C_{ie}(t) \leq{C_{ie,max}}$, and $\theta_{i}=V_ip_{e,max}+D_{ie,max} \leq B_{i,max}-C_{ie,max}$. Therefore, $B_{i}(t+1) \leq B_{i}(t)+C_{ie,max} < \theta_{i} + C_{ie,max} \leq B_{i,max}$.
\item Case 2: $B_{i}(t)\geq \theta_{i}$. In this case, $C_{ie}(t)=0$. The battery will not be charged at time slot $t$.  Therefore, $B_{i}(t+1) \leq B_{i}(t) \leq B_{i,max}$.

\item Case 3: $B_{i}(t)<D_{ie,max}$. In this case, $A_{i}(t)<D_{ie,max}-\theta_{i}=-V_ip_{e,max}$. Then, $A_{i}(t)+V_i\beta_{i}(t)<A_{i}(t)+V_ip_{e,max}<0$. According to (\ref{lg}) and (\ref{cde}), $D_{ie}(t)=0$. Therefore, $B_{i}(t+1) \geq B_{i}(t) \geq 0$.
\item Case 4: $B_{i}(t)\geq D_{ie,max}$. In this case, $B_{i}(t+1)= B_{i}(t)+C_{ie}(t)-D_{ie}(t)\geq B_{i}(t)-D_{ie}(t)\geq  0$.

\end{enumerate}

It is similar to analyze the bounds of $Y_{i}(t)$, $W_{i}(t)$, and $Y_{il}(t)$.
\section{Proof of Theorem 1  }
The optimal solution of problem (\ref{P2}) is obtained to minimize the drift-plus-penalty. Comparing this optimal solution with the result of the stationary random policy($\Pi$), the drift-plus-penalty term satisfies  

\begin{equation}
\begin{aligned}
&\Delta_{i}(t)+V_{i}\mathbb{E}\{C_{i}(t)\}
 \\ & \leq G_{i}+\mathbb{E}\{A_{i}(t)(C_{ie}^{*}(t)-D_{ie}^{*}(t))+F_{i}(t)(C_{iy}^{*}(t)-D_{iy}^{*}(t))\\&+Z_{i}(t)(C_{ih}^{*}(t)-D_{ih}^{*}(t))+\sum^{L_i}_{l=1}[I_{il}(t)(D_{iyl}^{*}(t)+d_{il}^{*}(t)\\&-Y_{ifl}^{*}(t)-h_{il}^{*}(t))]
+V_{i}(
p_{y}(t)d_{i}^{*}(t)
+E_{i}^{*}(t)p_{e}(t)\\&+ (P_{i}^{CHP,*}(t)+H_{i}^{CHP,*}(t)+H_{i}^{b,*}(t))p_{g}(t)+
\beta_{i}(t) X_{i}^{*}(t)\\&-\alpha_{i}(t) S_{i}^{*}(t)-E_{io}^{*}(t)p_{eo}(t))\}
 \\ & \leq G_{i}+\mathbb{E}\{A_{i}(t)(C_{ie}^{\Pi}(t)-D_{ie}^{\Pi}(t))+F_{i}(t)(C_{iy}^{\Pi}(t)-D_{iy}^{\Pi}(t))\\&+Z_{i}(t)(C_{ih}^{\Pi}(t)-D_{ih}^{\Pi}(t))+\sum^{L_i}_{l=1}[I_{il}(t)(D_{iyl}^{\Pi}(t)+d_{il}^{\Pi}(t)\\&-Y_{ifl}^{\Pi}(t)-h_{il}^{\Pi}(t))]
+V_{i}(
p_{y}(t)d_{i}^{\Pi}(t)
+E_{i}^{\Pi}(t)p_{e}(t)\\&+(P_{i}^{CHP,\Pi}(t)+H_{i}^{CHP,\Pi}(t)+H_{i}^{b,\Pi}(t))p_{g}(t)+\beta_{i}(t) X_{i}^{\Pi}(t)\\&-\alpha_{i}(t) S_{i}^{\Pi}(t)-E_{io}^{\Pi}(t)p_{eo}(t))\}
\end{aligned}
\end{equation}

According to (\ref{cd2}) and the stationary randomized policy that achieves the optimal cost $C_{ir}^{opt}$, the drift-plus-penalty term satisfies
%
\begin{equation}
\Delta_{i}(t)+V_{i}\mathbb{E}\{C_{i}(t)\} \leq G_{i} + V_{i}C_{ir}^{opt} \leq G_{i} + V_{i}C_{i}^{opt}
\end{equation}

Summing across $t \in \{1,2,...,T\}$, the sum term satisfies 
\begin{equation}
\mathbb{E}\{Q_{i}(T)-Q_{i}(1)\}+\sum_{t=1}^{T}V_{i}\mathbb{E}\{C_{i}(t)\} \leq TG_{i} + TV_{i}C_{i}^{opt}
\end{equation}

Dividing both sides by $TV_{i}$ and taking $T \rightarrow \infty$, the time-average cost term satisfies 
\begin{equation}
\lim_{T \rightarrow \infty} \frac{1}{T} \sum_{t=1}^{T} \mathbb{E}\{ C_{i}(t) \} \leq C_{i}^{opt} + \frac{G_{i}}{V_{i}}
\end{equation}

\end{appendix}

\balance
\section*{References}
\biboptions{square, numbers, sort&compress}
\bibliographystyle{elsarticle-num}
\bibliography{elsarticle-template}

\end{document}